\documentclass{elsart}
\usepackage{natbib}
\usepackage{amssymb}
\usepackage{graphicx}

\newcommand{\rd}{{\rm d}}
\newcommand{\ri}{{\rm i}}

\newcommand{\kL}{k_{\rm L}}
\newcommand{\Er}{E_{\rm r}}
\begin{document}

\begin{frontmatter}


\title{Kilohertz-driven Bose-Einstein condensates in optical lattices}

\author[label1,label2,label3]{Ennio Arimondo},
\author[label1,label2,label3]{Donatella Ciampini},
\author[label4]{Andr\'e Eckardt},
\author[label5]{Martin Holthaus},
\and
\author[label1,label2]{Oliver Morsch}

\address[label1]{Dipartimento di Fisica ``E.~Fermi'',
	Universit\`a di Pisa, Largo Pontecorvo 3, 56127 Pisa, Italy}
\address[label2]{CNR-INFM, Largo Pontecorvo 3, 56127 Pisa, Italy}
\address[label3]{CNISM UdR Universit\`a di Pisa, Largo Pontecorvo 3,
    	56127 Pisa, Italy}
\address[label4]{Max-Planck-Institut f\"ur Physik komplexer Systeme,
	N\"othnitzer Stra{\ss}e 38, D-01187 Dresden}
\address[label5]{Institut f\"ur Physik, Carl von Ossietzky Universit\"at,
	D-26111 Oldenburg, Germany}

\begin{abstract}
We analyze time-of-flight absorption images obtained with dilute Bose-Einstein
condensates released from shaken optical lattices, both theoretically and
experimentally. We argue that weakly interacting, ultracold quantum gases 
in kilohertz-driven optical potentials constitute equilibrium systems 
characterized by a steady-state distribution of Floquet-state occupation
numbers. Our experimental results consistently indicate that a driven ultracold
Bose gas tends to occupy a single Floquet state, just as it occupies a single
energy eigenstate when there is no forcing. When the driving amplitude is
sufficiently high, the Floquet state possessing the lowest mean energy does
not necessarily coincide with the Floquet state connected to the ground state 
of the undriven system. We observe strongly driven Bose gases to condense into
the former state under such conditions, thus providing nontrivial examples of 
dressed matter waves.
\end{abstract}


% 03.75.Lm :	Tunneling, Josephson effect, Bose-Einstein condensates in
%		periodic potentials, solitons, vortices, and topological
%		excitations.
% 67.85.De 	Dynamic properties of condensates; excitations,
%		and superfluid flow
% 67.85.Hj : 	Bose-Einstein condensates in optical potentials
% 05.30.Jp : 	Quantum statistical mechanics: Boson systems

\end{frontmatter}

% Main text
% =========

\section{Introduction}
\label{sec:S_1}

There is a growing interest in ultracold atoms confined in time-periodically 
driven optical lattices. It was pointed out already in 1997 that a
metal-insulator-like transition undergone by ultracold atoms in quasiperiodic
optical lattices can be induced by changing the amplitude of a time-periodic
driving force \citep{DreseHolthaus97}. But the field became truly active only
10 years later, after the unambiguous experimental observation of dynamical
tunneling suppression with Bose-Einstein condensates, and even of 
driving-induced reversal of the sign of the tunneling matrix element, by the 
Pisa group \citep{LignierEtAl07}, and after related single-particle experiments
performed with driven double-well potentials in Heidelberg \citep{KierigEtAl08}. 
Several groups then have monitored the dynamics under the combined action of
both a homogeneous time-independent and a time-periodic driving force, either
with non-condensed cold atoms \citep{IvanovEtAl08,AlbertiEtAl09}, or with 
Bose-Einstein condensates \citep{SiasEtAl08,HallerEtAl10}; such dynamics may   
be viewed as an analog of photon-assisted tunneling \citep{EckardtEtAl05a}.  
The Pisa group subsequently also reported the coherent control of the 
superfluid-to-Mott insulator transition in shaken three-dimensional optical 
lattices \citep{ZenesiniEtAl09}.
 
The recent detection of spontaneous breaking of time-reversal symmetry 
with fast-oscillating triangular optical lattices by \citet{StruckEtAl11} 
constitutes a further milestone in this line of research. Moreover, 
photon-assisted tunneling in a strongly correlated Bose gas has very recently 
been investigated, possibly allowing for applications to topological physics 
and quantum computing \citep{MaEtAl11}. Also active control of correlated 
tunneling in ac-driven optical superlattices has been observed, providing a 
novel approach to the realization of $XXZ$ spin models \citep{ChenEtAl11}. 

On the theoretical side, several studies indicate promising future perspectives:
\citet{EckardtHolthaus08} have suggested to perform avoided-level-crossing
spectroscopy with ultracold Bose gases in strongly driven optical lattices,
while \citet{EckardtEtAl10} have worked out a proposal for simulating 
frustrated quantum antiferromagnetism with oscillating triangular lattices.
Later \citet{TokunoGiamarchi11} have pointed out that a small periodic phase
modulation of an optical lattice can give direct access to the system's
conductivity. Furthermore, \citet{TsujiEtAl11} have argued that ac forcing
may even change the interparticle interaction from repulsive to attractive,
thereby enabeling one to simulate an effectively attractive Hubbard model with
a temperature below the superconducting transition temperature.

On a first superficial glance, ultracold atoms in shaken optical lattices,
with shaking frequencies on the order of a few to tens of kilohertz, might 
appear as typical examples of nonequilibrium systems. But this naive view is 
not correct. Quantum systems governed by a time-periodic Hamiltonian possess a 
particular basis, consisting of Floquet states, with respect to which one finds
occupation  numbers which remain {\em constant\/} in time, with the periodic
time-dependence already being incorporated into the basis states themselves.
Thus, a weakly interacting quantum gas subjected to time-periodic forcing,
and possibly exposed to some sort of noise, will be statistically characterized
by a time-independent distribution of Floquet-state occupation numbers. What
is that {\em equilibrium\/} distribution?

In the present tutorial article we develop this line of thought, and discuss
experimental data, obtained with Bose-Einstein condensates in kilohertz-shaken
optical lattices, which we interpret as evidence for Floquet-based dynamics.
We proceed as follows: In Sec.~\ref{sec:S_2} we review basic elements of
the Floquet picture, and speculate about the existence of analogs of the
Bose-Einstein and the Fermi-Dirac distribution function for isolated,
time-periodically driven quantum gases. We then turn in Sec.~\ref{sec:S_3} 
to the realization of periodically shaken optical lattices, 
extending the analysis of \citet{MadisonEtAl97}, and pay particular 
attention to the distinction between the co-moving frame of reference, in
which the trapped quantum gas experiences a spatially homogeneous inertial
force, and the laboratory frame, in which the measurements are performed.
In Sec.~\ref{sec:S_4} we briefly introduce the driven Bose-Hubbard model
\citep{EckardtEtAl05b}. In Sec.~\ref{sec:S_5} we calculate momentum
distributions expected in time-of-flight absorption imaging for matter waves
occupying a single Floquet state. We consider two particular situations: An
interacting gas periodically driven at high frequencies, and a noninteracting
gas subjected to a force with arbitrary time-dependence. Here we also point 
out certain characteristic differences between time-of-flight absorption 
images obtained from shaken lattices utilizing the inertial force, and the 
corresponding images acquired when the lattice remains at rest, while the force
is exerted, {\em e.g.\/}, through the harmonic modulation of a levitation
gradient \citep{HallerEtAl10}. In Sec.~\ref{sec:S_6} we tie the various strands
together and report our measurements, which suggest the existence of Floquet
condensates occupying the ``lowest'' available Floquet state. While we consider
only one comparatively simple experimental setting, it appears likely that
our findings are of a more general nature. In the final Sec.~\ref{sec:S_7}
we briefly spell out our main conclusions.

\section{The quest for Floquet condensates}
\label{sec:S_2}

The notion of Bose-Einstein condensation of an almost ideal gas
\citep{Einstein24,Einstein25,London38} appears to be inextricably linked
to the energy eigenstates of some single-particle Hamiltonian: When the
temperature of a gas of weakly interacting Bose particles becomes lower than
a certain critical temperature, the gas ``condenses'' into the single-particle
ground state, meaning that this ground state becomes macroscopically occupied.
It might seem that this concept cannot be applied to explicitly time-dependent
systems, for which, in general, there are no stationary states. Yet, there is
one notable exception: If a single-particle Hamiltonian $H^{(1)}(t)$ is
periodic in time, so that $H^{(1)}(t) = H^{(1)}(t+T)$ for some period $T$,
then Floquet's theorem \citep{Floquet83,Kuchment93} suggests the existence
of a set of distinguished solutions $|\psi_n(t)\rangle$, known as Floquet
states, to the time-dependent Schr\"odinger equation
$\ri\hbar\partial_t |\psi(t)\rangle = H^{(1)}(t) |\psi(t)\rangle$;
these Floquet states are in many ways analogous to the usual energy
eigenstates of time-independent Hamiltonian operators
\citep{Shirley65,Zeldovich67,Ritus67,Sambe73,BreuerHolthaus91,ChuTelnov04}.
They have the particular form
\begin{equation}
	|\psi_n(t)\rangle = | u_n(t) \rangle
	\exp(-\ri \varepsilon_n t/\hbar) \; ,
\label{eq:FLS}
\end{equation}
where the functions $|u_n(t)\rangle = |u_n(t+T)\rangle$ inherit the period~$T$
of the underlying Hamiltonian; the index $n$ denotes a set of quantum numbers
specifying the state. The quantities $\varepsilon_n$, which determine the phase
factors accompanying the time-evolution of the Floquet states~(\ref{eq:FLS})
in the same manner as energy eigenvalues of time-independent systems determine
the phase factors expressing the evolution of their energy eigenstates, are
dubbed quasienergies \citep{Zeldovich67,Ritus67}.

There is a compelling reason for considering these Floquet states. Namely, at
each instant $t_0$ the set of Floquet functions  $\{ |u_n(t_0)\rangle \}$ is
complete in the physical Hilbert space on which $H^{(1)}(t)$ acts. Hence,
{\em any\/} solution $|\psi(t)\rangle$ to the time-dependent Schr\"odinger
equation admits an expansion of the form
\begin{equation}
	|\psi(t)\rangle = \sum_n c_n | u_n(t) \rangle
	\exp(-\ri \varepsilon_n t/\hbar)
\label{eq:FSE}
\end{equation}
with {\em time-independent\/} coefficients $c_n$. Therefore, under conditions
of perfectly coherent evolution the occupation probabilities $|c_n|^2$ of the
Floquet states remain constant, despite the periodic time dependence of the
Hamiltonian. These coefficients then keep a memory of how the system has been
prepared, that is, how the time-periodic forcing has been turned on in the
past, whereas the Floquet states and their quasienergies are independent of
this history.

If, however, the evolution is not coherent, but the system is in contact with
some sort of thermal environment or heat bath, what would be the equilibrium
distribution of the Floquet-state occupation probabilities which establishes
itself in the long run, irrespective of the particular initial condition?
An important step towards a general answer to this question has been taken
by \citet{BreuerEtAl00}, who have studied anharmonic oscillators subjected to
strong time-periodic forcing while being weakly coupled to thermal degrees of 
freedom provided by a surrounding, and who have shown that the corresponding
quasi\-stationary density matrix actually is diagonal in the Floquet
representation. This line of investigation has recently been taken up by
\citet{KetzmerickWustmann10}, with a detailed view on the classical-quantum
correspondence, after general aspects of such ``periodic thermodynamics''
had been discussed by \citet{Kohn01}.
 
Turning now from the single-particle dynamics to that of an ideal gas of
identical quantum particles subjected to some periodically time-dependent
external influence, such as a dilute gas of Bose particles stored in a
confining potential which is modulated periodically in time, an even more
intriguing question poses itself: What are the statistics of the Floquet-state
occupation numbers? In other words, what is the replacement for the familiar
Bose-Einstein (or Fermi-Dirac, when dealing with Fermions) distribution
function in a periodically time-dependent setting?

In order to appreciate what lies behind this question, let us briefly retrace
the steps which led \citet{Einstein25} to what is now known as Bose-Einstein
statistics, by now having become basic knowledge of statistical physics
\citep{Pathria96}: Consider an isolated ideal Bose gas consisting of $N \gg 1$
particles possessing a total energy $E_{\rm tot}$, being kept in some large
confinement. In view of the corresponding denseness of the single-particle
spectrum we divide the energy axis into small intervals, the $i$th such
interval containing $g_i$ single-particle states with practically identical
energy $E_i$. We then denote the number of particles occupying the $g_i$ states
in the $i$th interval as $n_i$, and ask for the most probable set $\{ n_i^* \}$
of these cell occupation numbers. All admissible sets $\{ n_i \}$ obviously
have to comply with the two constraints
\begin{equation}
	\sum_i n_i = N
\label{eq:PNC}
\end{equation}
and
\begin{equation}
	\sum_i n_i E_i = E_{\rm tot} \; ;
\label{eq:ENC}	
\end{equation}
the latter holding approximately, reflecting the above coarse-graining
procedure, but to good accuracy. The number of microstates associated with
a given set $\{ n_i \}$ is
\begin{equation}
	\Omega[\{n_i\}] = \prod_i {n_i + g_i - 1 \choose n_i} \; ,
\end{equation}
because the binomial coefficients quantify the number of possibilities to
distribute $n_i$ indistinguishable particles over $g_i$ states ``with
repetition'', as is characteristic for Bosons. Assuming that both $n_i \gg 1$
and $g_i \gg 1$, and employing Stirling's formula, one then obtains the
entropy
\begin{equation}
	\ln \Omega[\{n_i\}] = \sum_i \left[
	  n_i \ln\left( 1 + \frac{g_i}{n_i}\right)
	+ g_i \ln\left( \frac{n_i}{g_i} + 1 \right) \right] \; .
\label{eq:ENT}
\end{equation}
The searched-for most probable occupation numbers $\{ n_i^* \}$ maximize
this expression, subject to the two constraints~(\ref{eq:PNC}) and
(\ref{eq:ENC}). Introducing a Lagrangian multiplier~$\alpha$ to account for
the conservation of the particle number, and a further multiplier~$\beta$
to account for the conservation of energy, we are thus led to the variational
problem
\begin{equation}
	\delta \left( \ln \Omega[\{n_i\}] - \alpha \sum_i n_i
	-\beta \sum_i n_i E_i  \right)_{n_i = n_i^*} = 0 \; .
\label{eq:EVA}
\end{equation}
After brief calculation, this gives
\begin{equation}
	\sum_i \left[ \ln\left(1 + \frac{g_i}{n_i}\right)
	- \alpha -\beta E_i \right]_{n_i = n_i^*} \delta n_i = 0 \;
\end{equation}
and thus results in the Bose-Einstein distribution
\begin{equation}
	\frac{n_i^*}{g_i} = \frac{1}{\exp(\beta E_i + \alpha) - 1} \; ,
\end{equation}
quantifying the most probable occupation number of an individual
single-particle state with energy $E_i$ under microcanonical conditions.
Finally, the multiplier $\beta$ is identified with $1/(k_{\rm B}T)$, with
$k_{\rm B}$ denoting Boltzmann's constant and $T$ being the temperature of
the gas, while $\alpha = -\mu/(k_{\rm B}T)$ is related to its chemical
potential~$\mu$ \citep{Einstein25,Pathria96}.

When trying to adapt this reasoning to time-periodically driven quantum gases,
and to determine the corresponding Floquet-state occupation numbers, we can
immediately carry over the first constraint~(\ref{eq:PNC}): The sum of all
occupation numbers has to equal the total number of particles. This may seem
trivial, but it is not, because already here we are making essential use
of Floquet theory: It is only because the coefficients $c_n$ in the
expansion~(\ref{eq:FSE}) are time-independent that we can assign occupation
numbers to the Floquet states; only then can one ask for the associated
distribution function.

It is, however, not obvious what becomes of the second
constraint~(\ref{eq:ENC}), because energy eigenvalues do not exist for a
time-periodic system, and the quasienergies of the Floquet states cannot be
considered as their proper substitutes in this context. To understand the
quasienergy concept more deeply, we insert the Floquet states~(\ref{eq:FLS})
into the time-dependent Schr\"odinger equation, and deduce
\begin{equation}
	\left( H^{(1)}(t) - \ri\hbar\frac{\partial}{\partial t} \right)
	| u_n (t) \rangle \! \rangle
	= \varepsilon_n | u_n (t) \rangle \! \rangle \; .
\label{eq:EVE}
\end{equation}	
The conspicuous notation employed here, {\em i.e.\/}, writing the Floquet
functions $| u_n (t) \rangle \! \rangle$ with a double right angle instead of
$| u_n (t) \rangle$ as in Eq.~(\ref{eq:FLS}), has a deep significance. Namely,
Eq.~(\ref{eq:EVE}) has to be regarded as an {\em eigenvalue equation\/} for
the quasienergies; within the Floquet framework, this eigenvalue equation
takes over the role played for time-independent systems by the stationary
Schr\"odinger equation. Unlike the latter, this quasienergy eigenvalue
equation~(\ref{eq:EVE}) lives in an {\em extended Hilbert space\/} of
$T$-periodic functions \citep{Sambe73}. In contrast to the usual physical
setting, in which ``time'' emerges as an evolution variable, in that
extended space time plays the role of a {\em coordinate\/}, and therefore
needs to be integrated over in the associated scalar product: Denoting the
momentary scalar product of two $T$-periodic functions $|u_1(t)\rangle$ and
$|u_2(t)\rangle$ in the physical space as $\langle u_1(t) | u_2(t) \rangle$,
their scalar product in the extended space reads \citep{Sambe73}
\begin{equation}
	\langle \! \langle u_1 | u_2 \rangle \! \rangle \equiv
	\frac{1}{T} \int_0^T \! \rd t \, \langle u_1(t) | u_2(t) \rangle \; .
\label{eq:SCP}
\end{equation}
Thus, we write $| u_n (t) \rangle$ when considering a Floquet function in the
physical Hilbert space, as in Eqs.~(\ref{eq:FLS}) and (\ref{eq:FSE}), whereas
$| u_n (t) \rangle \! \rangle$ refers to that same function when viewed as an
element of the extended space, as in Eq.~(\ref{eq:EVE}).

This apparently formal observation has an important, physically meaningful
consequence. Suppose that $| u_n (t) \rangle \! \rangle$ is a solution to the
eigenvalue equation~(\ref{eq:EVE}) with quasienergy $\varepsilon_n$, define
$\omega = 2\pi/T$, and let $m$ be an arbitrary (positive, zero, or negative)
integer. Then $| u_n (t) \exp(\ri m \omega t) \rangle \! \rangle$ also is
a $T$-periodic solution, with quasienergy $\varepsilon_n + m\hbar\omega$.
For $m \neq 0$ these two Floquet functions obviously are orthogonal with
respect to the scalar product~(\ref{eq:SCP}), and represent {\em different\/}
eigenfunctions in the extended space. On the other hand, when returning to the
physical space and forming the actual Floquet states~(\ref{eq:FLS}), one has
\begin{equation}
	| u_n(t) \exp(\ri m \omega t) \rangle
	\exp\!\Big(-\ri(\varepsilon_n + m\hbar\omega)t/\hbar\Big)
	= | u_n(t) \rangle \exp(-\ri \varepsilon_n t/\hbar) \; ,  	
\end{equation}
so that both solutions represent {\em the same\/} physical state. Hence, a
Floquet state does not correspond to a single solution to Eq.~(\ref{eq:EVE}),
but rather to a whole class of such solutions, labeled by the index~$n$,
while individual representatives of such a class are distinguished by the
integer~$m$. This subtlety already reflects itself in the
expansion~(\ref{eq:FSE}): Although {\em all\/} functions
$\{ |u_n(t) \exp(i m \omega t) \rangle\!\rangle \}$, with $n$ ranging over
all state labels and $m$ extending over all integers, are required for the
completeness relation in the extended space, {\em only one\/} representative
from each class is required in Eq.~(\ref{eq:FSE}), where no sum over the
``photon index'' $m$ appears. By the same token, the quasienergy of a physical
Floquet state is determined only up to an integer multiple of the ``photon''
energy $\hbar\omega$. In accordance with the analogous terminology used in
solid-state physics, the quasienergy spectrum is said to consist of an infinite
set of identical ``Brillouin zones'' of width $\hbar\omega$, covering the
entire energy axis, each state placing one of its quasienergy representatives
in each zone. This means, in particular, that there is no natural ``quasienergy
ordering'': Without additional specification it is meaningless to ask whether
one given Floquet state lies ``below'' another. And there is still a further
complication of a more mathematical nature: Because one finds one quasienergy
representative of each state in each Brillouin zone, the quasienergy spectrum
is ``dense'', and it may be technically difficult to decide whether one has
a dense pure point spectrum, so that the expansion~(\ref{eq:FSE}) can be taken
literally and the system is stable, possessing a quasiperiodic wave function,
or whether there is an absolutely continuous spectrum, allowing for diffusive
energy growth \citep{BunimovichEtAl91,Howland92a}. We bypass this problem by
restricting ourselves to stable systems with a pure point spectrum, such as
the forced anharmonic oscillators with superquadratic potentials investigated
by \citet{Howland92b}.

These considerations clearly reveal that, although there are Floquet-state
occupation numbers, the required analog of the energy constraint~(\ref{eq:ENC})
for time-periodically forced microcanonical close-to-ideal Bose gases cannot
involve quasienergies directly. However, each Floquet state possesses a
mean energy \citep{FainshteinEtAl78}
\begin{equation}	
	\overline{E}_n \equiv \frac{1}{T} \int_0^T \! \rd t \,
	\langle u_n(t) | H^{(1)}(t) | u_n(t) \rangle
	= \langle \! \langle u_n | H^{(1)} | u_n \rangle \! \rangle
\label{eq:FME}
\end{equation}
which obviously is independent of the choice of the representative
$| u_n \rangle \! \rangle$. If we now repeat the coarse-graining procedure
employed in the formulation of the energy constraint~(\ref{eq:ENC}),
that is, if we divide the energy axis into small cells, the $i$th cell
now containing $g_i$ Floquet states having almost the same mean energy
$\overline{E}_i$, and if $\{ n_i \}$ denotes the corresponding sets of
cell occupation numbers, then it seems reasonable to demand
\begin{equation}
	\sum_i n_i \overline{E}_i = {\rm const.}
\label{eq:MEC}
\end{equation}
for an isolated system, thus obtaining a constraint which effectively may
replace the previous Eq.~(\ref{eq:ENC}). Here we assume that each driven 
particle sees the other driven particles as its heat bath, and undergoes 
subsequent short relaxation events only on a time scale which is large compared
to the period~$T$. In general, the justification of this constraint may require
the consideration of particular physical setups, and some more specifications 
may be needed. But if we tentatively accept this constraint~(\ref{eq:MEC}) 
as it stands, it is immediately clear how to proceed: We assign the 
entropy~(\ref{eq:ENT}) to a time-periodically forced isolated Bose gas with 
Floquet-state occupation numbers $\{ n_i \}$, account for the particle number 
conservation~(\ref{eq:PNC}) again by introducing a Lagrangian multiplier 
$\alpha$, and incorporate the conservation of the mean energy~(\ref{eq:MEC}) 
with the help of a further Lagrangian multiplier~$\gamma$. Then the most 
probable set of Floquet-state occupation numbers $\{ n_i^* \}$ is determined 
by the variational equation
\begin{equation}
	\delta \left( \ln \Omega[\{n_i\}] - \alpha \sum_i n_i
	-\gamma \sum_i n_i \overline{E}_i  \right)_{n_i = n_i^*} = 0 \; .
\end{equation}
Because this problem exactly parallels the problem~(\ref{eq:EVA}) already
considered by \citet{Einstein25}, we can directly take over the solution:
Under the above conditions, the most probable set of Floquet-state occupation
numbers is
\begin{equation}
	\frac{n_i^*}{g_i} =
	\frac{1}{\exp(\gamma \overline{E}_i + \alpha) - 1} \; ,
\end{equation}
suggesting that
\begin{equation}
	f(\overline{E}_n; \alpha, \gamma)
	= \frac{1}{\exp(\gamma \overline{E}_n + \alpha) - 1}
\label{eq:BED}
\end{equation}
is the expected occupation number of an individual Floquet state with mean
energy $\overline{E}_n$. Of course, this reasoning can easily be adpated
to Fermions, leading to the familiar ``plus'' sign in the denominator
\citep{Pathria96}.

These deliberations may have some profound physical implications. To begin
with, a time-periodically forced, stable Bose gas manifestly does {\em not\/}
constitute a nonequilibrium system, as one might naively assume, but rather
an equilibrium one, with equilibrium parameters $\alpha$ and $\gamma$. 
Most noteworthy, these parameters depend on the form of the periodic 
time-dependence, because so do the Floquet states, and hence their mean
energies~(\ref{eq:FME}), which enter into the modified
constraint~(\ref{eq:MEC}). Therefore, when writing
\begin{equation}
	\gamma = \frac{1}{k_{\rm B}\Theta}
\label{eq:DTH}
\end{equation}
in analogy to the familiar relation $\beta = 1/(k_{\rm B}T)$, the parameter
$\Theta$ introduced here is a temperature-like quantity which can be varied 
by changing, {\em e.g.\/}, the strength of an external time-periodic force; 
in general, $\Theta$ will differ from the usual temperature $T$ the gas would 
have if there were no such forcing. Likewise, when setting
\begin{equation}
	\alpha = -\frac{\nu}{k_{\rm B}\Theta} \; ,
\label{eq:DNU}
\end{equation}
$\nu$ is the corresponding, forcing-dependent chemical potential.

It follows that there are {\em Floquet condensates\/}: If the
$\Theta$-temperature is sufficiently low, a time-periodically forced
close-to-ideal Bose gas condenses into the single-particle Floquet state
possessing the lowest mean energy. This deduction immediately suggests a
further possibility: Suppose that the Hamiltonian under consideration has
the natural form $H^{(1)}(t) = H_0 + H_1(t)$, where $H_1(t)$ represents a
time-periodic force acting on the unperturbed system $H_0$ with adjustable
strength. In the regime of perturbatively weak forcing the mean-energy
ordering of the Floquet states of $H^{(1)}(t)$ is likely to be the same as
the energy ordering of the eigenstates of $H_0$ from which they have developed,
but in the nonperturbative regime this is no longer guaranteed; here the
Floquet state possessing the lowest mean energy not necessarily is connected
to the $H_0$-ground state. Under such conditions the Floquet state into which
an ultracold, time-periodically forced Bose gas condenses is determined by
the strength of the forcing, and may change when the latter is varied.

We close this section by emphasizing that the above reasoning hinges on 
two key issues: The assignment of occupation numbers to the Floquet states, 
which is borne out by theory, and the adoption of the mean-energy 
constraint~(\ref{eq:MEC}) as a building principle for a microcanonical Floquet 
ensemble, which may require further thoughts. Important pieces of evidence 
supporting our theoretical deductions are provided by the experimental 
observations reported in Sec.~\ref{sec:S_6} below. In order to facilitate 
the interpretation of our measurements, we first lay out some important 
ingredients in Secs.~\ref{sec:S_3} to \ref{sec:S_5}.

\section{The experimental setup: Shaken optical lattices}
\label{sec:S_3}

A one-dimensional (1D) optical lattice potential, which we write in the
form
\begin{equation}
	V(x) = \frac{V_0}{2}\cos(2\kL x) \; ,
\label{eq:OCL}	
\end{equation}
is generated by two linearly polarized counterpropagating beams of laser 
radiation with wave number~$\kL$, its depth $V_0$ being proportional to the 
laser intensity \citep{MorschOberthaler06}. Here the lattice coordinate~$x$ 
refers to the laboratory frame of reference. Considering an atom of mass~$M$ 
moving in this potential, the lattice depth~$V_0$ is measured in units of its
single-photon recoil energy~$\Er$,
\begin{equation}
	\Er = \frac{\hbar^2 \kL^2}{2M} \; ;
\end{equation}
typical optical lattices are up to tens of recoil energies deep. For example, 
when working with $^{87}$Rb in an optical lattice made up from laser light 
with wavelength $\lambda = 2\pi/\kL = 852$~nm one obtains
$\Er = 1.31 \cdot 10^{-11}$~eV. Thus, the typical depth of an optical lattice
is 11 orders of magnitude lower than that of the lattices encountered in
traditional solid-state physics. This implies that the characteristic
frequencies of ultracold atoms in optical lattices fall into the lower
kilohertz regime.

The pairs of counterpropagating beams needed to create optical lattices in one,
two, or three dimensions can either be realized by splitting a laser beam in 
two, or by retro-reflecting a beam off a mirror. In the former case, a small 
frequency difference $\Delta \nu(t)$ introduced between the two splitted laser 
beams with the help of acousto-optic modulators makes the lattice move with
velocity $v(t) = \Delta \nu(t) \lambda/2$ \citep{NiuEtAl96,BenDahanEtAl96,
MadisonEtAl98,LignierEtAl07,SiasEtAl08}. Therefore, in the laboratory frame
of reference an atom then experiences the potential
\begin{equation}
	V_{\rm lab}(x,t) =
	\frac{V_0}{2}\cos\!\Big(2\kL \left[ x - X_0(t) \right]\Big) \; ,
\label{eq:VLB}
\end{equation}
where
\begin{equation}
	X_0(t) = \frac{\lambda}{2}\int_{0}^t \! \rd \tau \, \Delta \nu(\tau)
	\; .
\label{eq:SHI}
\end{equation}
When using retro-reflected beams, lattice motion can be achieved by mounting
the mirror on a piezo-electric actuator, so that it shifts according to some 
prescribed protocol $X_0(t)$ \citep{ZenesiniEtAl09,AlbertiEtAl09}. As the 
shift range of such devices is in the micrometer regime, this method is 
suitable for inducing an oscillatory motion as needed for shaking lattices, 
but not for applying a constant acceleration.

In the following steps, the Schr\"odinger equation for the translational
motion of an atom in the shifted lattice, governed by the Hamiltonian
\begin{equation}
	H_{\rm lab}(t) = \frac{p^2}{2M} + V_{\rm lab}(x,t) \; ,
\end{equation}
is subjected to a unitary transformation to the frame of reference co-moving
with the lattice. First, the corresponding shift in position is implemented
by means of the unitary operator
\begin{equation}
	U_1 = \exp\!\left(\frac{\ri}{\hbar} X_0(t) \, p \right) \; ,
\end{equation}
implying
\begin{equation}
	U_1 x \, U_1^{\dagger} = x + X_0(t)
\end{equation}
and
\begin{equation}
	U_1 \left( -\ri\hbar \frac{\partial}{\partial t} \right) U_1^\dagger
	= -\ri\hbar \frac{\partial}{\partial t}
	- \dot{X}_0(t) \, p \; ,
\end{equation}
thus leading to the new Hamiltonian
\begin{equation}
	\widetilde{H}(t) =
	\frac{1}{2M}\left( p - M \dot{X}_0(t) \right)^2
	+ \frac{V_0}{2}\cos(2 \kL x)
	- \frac{M}{2} \dot{X}_0(t)^2 \; .
\end{equation}
Next, the shift of momentum showing up here is compensated through
\begin{equation}
	U_2 = \exp\!\left(-\frac{\ri}{\hbar} M \dot{X}_0(t) \, x \right) \; ,
\end{equation}
giving
\begin{equation}
	U_2 p \, U_2^\dagger = p + M \dot{X}_0(t)
\label{eq:TRP}
\end{equation}		
together with
\begin{equation}
	U_2 \left( -\ri\hbar \frac{\partial}{\partial t} \right) U_2^\dagger
	= -\ri\hbar \frac{\partial}{\partial t}
	+ M \ddot{X}_0(t) \, x \; .
\end{equation}
Finally, the merely time-dependent energy shift appearing in
$\widetilde{H}(t)$ is removed by the operator
\begin{equation}
	U_3 = \exp\!\left(-\frac{\ri}{\hbar} \frac{M}{2}
	\int_0^t \! \rd \tau \, \dot{X}_0^2(\tau) \right) \; ,
\end{equation}
since
\begin{equation}
	U_3 \left( -\ri\hbar \frac{\partial}{\partial t} \right) U_3^\dagger
	= -\ri\hbar \frac{\partial}{\partial t}
	+ \frac{M}{2} \dot{X}_0^2(t) \; .
\end{equation}
The full transformation to the co-moving frame is then realized by the
combined operation $U = U_3 U_2 U_1$, giving
\begin{equation}
	U \Big( H_{\rm lab}(t) - \ri \hbar \partial_t \Big) U^\dagger =
	H_0 + H_1(t) - \ri\hbar\partial_t \; ,
\end{equation}
with
\begin{equation}
	H_0 = \frac{p^2}{2M} + \frac{V_0}{2}\cos(2\kL x)
\end{equation}
denoting the single-particle Hamiltonian pertaining to the undriven lattice,
and
\begin{equation}
	H_1(t) = -F(t) x
\label{eq:SPF}			
\end{equation}
introducing a homogeneous inertial force $F(t)$ acting in the co-moving
frame, determined by the lattice motion according to
\begin{equation}
	F(t) = - M \ddot{X}_0(t) \; .
\label{eq:HIF}
\end{equation}
In particular, when a purely sinusoidal frequency shift with an initial
phase~$\phi$ is suddenly applied at $t = 0$,
\begin{equation}
	\Delta \nu (t) = \left\{ \begin{array}{ll}
	0 & \; , \quad t < 0 \; , \\
	\Delta \nu_{\rm max} \sin(\omega t + \phi) & \; , \quad t > 0 \; ,
		\end{array} \right.
\label{eq:PHI}
\end{equation}	
Eqs.~(\ref{eq:VLB}) and (\ref{eq:SHI}) give
\begin{equation}
	V_{\rm lab}(x,t) = \frac{V_0}{2}
	\cos\!\Big(2 \kL [x + L\{\cos(\omega t + \phi) - \cos(\phi)\}]\Big)
\label{eq:OSL}
\end{equation}
for $t > 0$, so that the lattice is shaken in the laboratory frame with
the amplitude
\begin{equation}
	L = \frac{\lambda \Delta \nu_{\rm max}}{2\omega} \; .
\label{eq:AMP}
\end{equation}
According to Eq.~(\ref{eq:SHI}), one has
$\dot{X}_0(t) = \lambda \Delta\nu(t)/2$.
Therefore, this protocol~(\ref{eq:PHI}) effectuates a sudden jump of the
lattice's velocity at time $t = 0$, unless $\phi = 0$ or $\phi = \pi$. By 
virtue of Eq.~(\ref{eq:HIF}), the force felt by the atoms in the co-moving 
frame of reference then is composed of a monochromatic oscillating drive 
acting for $t > 0$, and of a delta-like kick acting at the moment of turn-on: 
Formally, one obtains
\begin{equation}
	F(t) =  -F_1 \cos(\omega t + \phi) \, \Theta(t)
	        -\frac{F_1}{\omega} \sin(\phi) \, \delta(t) \; ,		
\label{eq:MOF}
\end{equation}
where $\Theta(t)$ denotes the Heaviside function, and the driving amplitude
is given by
\begin{equation}
	F_1 = M \frac{\lambda}{2} \Delta \nu_{\rm max} \omega
	= M L \omega^2 \; .
\label{eq:MAF}	
\end{equation}
Needless to say, in reality the delta-kick experienced at $t = 0$ has a finite
sharpness, determined by the short-time details of the actual velocity jump.
Nonetheless, this kick has an important experimental consequence, as will be
discussed later.

As a dimensonless measure of the shaking or driving strength we introduce the 
quantity
\begin{equation}
	K_0 = \frac{F_1d}{\hbar\omega}
	= \frac{\pi^2}{2} \frac{\omega}{\omega_{\rm r}} \frac{L}{d} \; ,
\end{equation}	
where $d = \lambda/2$ is the lattice constant. Thus, with driving
frequencies on the order of the recoil frequency $\omega_{\rm r} = \Er/\hbar$,
and modulation amplitudes~(\ref{eq:AMP}) on the order of the lattice constant,
one can easily reach the nonperturbative regime $K_0 > 1$. It would be quite
hard to realize corresponding conditions in laser-irradiated traditional 
solids without introducing, {\em e.g.\/}, additional polarization effects, 
or even damaging the sample. Therefore, strongly shaken optical lattices may 
also be viewed as ``strong-field simulators'' which allow one to study even
superstrong-field-induced multiphoton-like processes in periodic potentials,
such as interband transitions, in their purest form
\citep{ArlinghausHolthaus10}.

\section{The driven Bose-Hubbard model}
\label{sec:S_4}

\begin{figure}[t]
\centerline{\includegraphics[angle=-90., width = 0.7\linewidth]
	{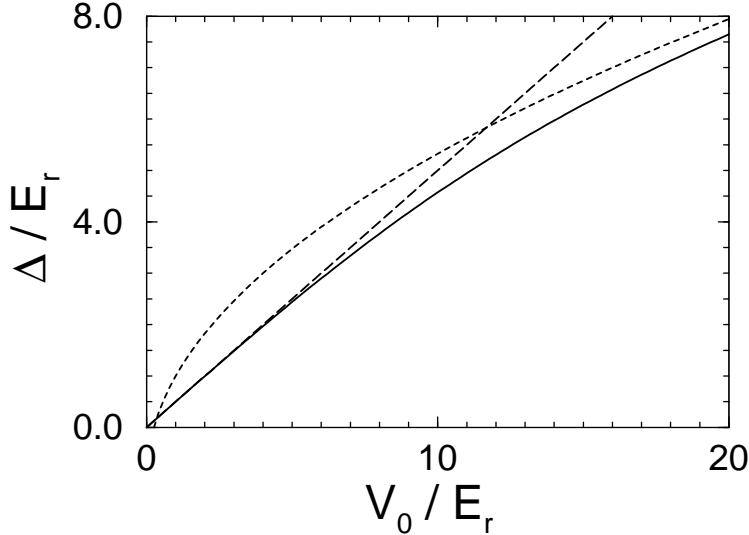}}
\caption[Fig.~1]{Exact energy gap $\Delta$ between the lowest two bands of
	an optical cosine lattice~(\ref{eq:OCL}), measured in multiples
	of the recoil energy~$\Er$ (full line), in comparison with the
	shallow-lattice estimate~(\ref{eq:SLE}) (long dashes), and
	with the deep-lattice estimate~(\ref{eq:DLE}) (short dashes).}
\label{F_1}
\end{figure}

In principle, the time-periodic motion~(\ref{eq:OSL}) always induces
transitions between the unperturbed energy bands of the optical lattice.
However, if the energy scale $\hbar\omega$ associated with the modulation
frequency remains small compared to the energy gap~$\Delta$ between the lowest
two bands, and if the driving amplitude remains sufficiently low, the dynamics
of driven  ultracold atoms remain restricted to the lowest band at least to
good approximation. Since the band structure of a cosine lattice~(\ref{eq:OCL})
is determined by the characteristic values of the Mathieu equation
\citep{Slater52}, the known expansions of these values
\citep{AbramowitzStegun65} can be employed for obtaining estimates of the
gap width, resulting in
\begin{equation}
	\Delta / \Er \approx V_0 / (2\Er)
\label{eq:SLE}
\end{equation}
for shallow lattices, and
\begin{equation}
	\Delta / \Er \approx 2\sqrt{V_0/\Er} - 1
\label{eq:DLE}
\end{equation}
for fairly deep ones. Fig.~\ref{F_1} demonstrates that these two
approximations indeed provide a reasonable estimate of the exact
band gap for all $V_0/\Er$, if one switches from the shallow-lattice
result~(\ref{eq:SLE}) to the deep-lattice formula~(\ref{eq:DLE}) at
$V_0/\Er \approx 11.7$. As a figure of merit, the exact gap width is
$\Delta = 4.572 \, \Er$ for a lattice with depth $V_0 = 10 \, \Er$.

In our experiments \citep{LignierEtAl07,SiasEtAl08,EckardtEtAl09,ZenesiniEtAl09}
we work with Bose-Einstein condensates of $^{87}$Rb consisting of about
$5 \times 10^4$ atoms in a shaken 1D optical lattice ($\lambda = 852$~nm or
$842$~nm) with depths $V_0/\Er$ ranging from 3 to 10, employing scaled shaking
frequencies $\hbar\omega/\Er$ between roughly 0.1 and~2 (corresponding to
$\omega/(2\pi)$ between 0.3 and 6~kHz). Under such conditions the single-band
approximation is viable, provided the forcing is not too strong, since
crossing the gap would require higher-order multiphoton-like transitions
\citep{ArlinghausHolthaus10}. In order to model these driven, interacting
many-body systems we resort to the co-moving frame of reference, and then
follow the standard route \citep{JakschEtAl98,JakschZoller05,BlochEtAl08}.
Employing a basis of site-localized Wannier functions pertaining to the lowest
Bloch band, and assuming a reasonable lattice depth $V_0/\Er \gg 1$, it
suffices to retain only the hopping matrix element~$J$ connecting neighboring
sites. The accuracy of this approximation has been assessed quantitatively
by \citet{BoersEtAl07} and by \citet{EckardtEtAl09}; for instance, when
$V_0/\Er = 10$ the neglected matrix element connecting next-to-nearest
neighbors is actually smaller than~$J$ by a factor of about $0.012$. Moreover,
the van der Waals length of alkali atoms typically amounts to just a few 
nanometers \citep{BlochEtAl08} and thus is significantly smaller than the 
lattice constant $d = \lambda/2$, so that only the on-site interaction among 
the atoms has to be accounted for; this is done in terms of a parameter~$U$ 
which quantifies the interaction energy of one pair of atoms occupying the same
lattice site~\citep{SchneiderEtAl09}. In short, when adopting these three 
approximations (single-band, nearest-neighbor, and on-site), an ultracold gas 
of bosonic atoms in a time-periodically shaken optical lattice is described 
in the co-moving frame by the explicitly time-dependent many-body
Hamiltonian\footnote{We use the ``hat''-symbol to indicate operators acting on
the bosonic Fock space} \citep{EckardtEtAl05b,CreffieldMonteiro06}
\begin{equation}
	\hat{H}(t) = \hat{H}_0 + \hat{H}_1(t) \; ,
\label{eq:DBH}
\end{equation}
where	
\begin{equation}
	\hat{H}_0 = -J \sum_\ell\left(
	\hat{b}^\dagger_\ell \hat{b}^{\phantom\dagger}_{\ell+1} +
	\hat{b}^\dagger_{\ell+1} \hat{b}^{\phantom\dagger}_\ell \right)
	+ \frac{U}{2}\sum_\ell \hat{n}_\ell \left( \hat{n}_\ell - 1 \right)
\label{eq:BHH}
\end{equation}
is the standard Bose-Hubbard model for a 1D lattice
\citep{FisherEtAl89,JakschEtAl98,JakschZoller05}, with
$\hat{b}^{\phantom\dagger}_\ell$ denoting the bosonic annihilation operator
for atoms occupying the Wannier state at the $\ell$th site;
$\hat{n}_\ell = \hat{b}^\dagger_\ell \hat{b}^{\phantom\dagger}_\ell$
is the number operator for that site. Recalling that the position
operator~$x$ mediating the inertial force built into the single-particle
Hamiltonian~(\ref{eq:SPF}) translates, on the many-body level, into
$d\sum_\ell \ell \hat{n}_\ell$, with $d = \lambda/2$ for the distance
between two sites, the time-periodic forcing now is introduced through
\begin{equation}	
	\hat{H}_1(t) =	K \cos(\omega t) \sum_\ell \ell \hat{n}_\ell \; ,
\label{eq:DCF}
\end{equation}
where the strength~$K$ is related to the forcing amplitude~(\ref{eq:MAF})
by $K = F_1 d$. Here we disregard an initial phase $\phi$, which would refer
to a sudden turn-on of the force, as in Eq.~(\ref{eq:PHI}). This driven
1D Bose-Hubbard model~(\ref{eq:DBH}) may be extended to higher spatial
dimensions \citep{ZenesiniEtAl09}; we remark that approximate expressions
relating the parameters $J$ and $U$ to the lattice depth have been provided
by \citet{Zwerger03} and by \citet{BlochEtAl08}.

\section{Interference patterns produced by Floquet states}
\label{sec:S_5}

It is essential to observe that the existence of Floquet states~(\ref{eq:FLS})
hinges solely on the periodicity of the given Hamiltonian in time. Thus,
besides the single-particle Floquet states which form the basis of the
statistical considerations in Sec.~\ref{sec:S_2}, there also are Floquet
states for periodically time-dependent, interacting many-body systems, such
as periodically driven Bose gases, which incorporate both the periodic
time-dependence and all interaction effects.

\subsection{Signatures of interacting shaken Bose gases}

Here we are concerned, in particular, with the Floquet states of the driven
Bose-Hubbard model~(\ref{eq:DBH}), and their experimental signatures. In order
to solve the quasienergy eigenvalue problem~(\ref{eq:EVE}) for this model,
the quasienergy operator $\hat{H}(t) - \ri \hbar \partial_t$ has to be
diagonalized in the associated extended Hilbert space \citep{Sambe73}, as
reviewed in Sec.~\ref{sec:S_2}. To this end, let $\{ n_\ell \}$ denote an
admissible set of site-occupation numbers $n_\ell$. The physical many-body
Hilbert space then is spanned by the set of all Fock states
\begin{equation}
	| \{ n_\ell \} \rangle =
	\prod_\ell \frac{(\hat{b}^\dagger_{\ell})^{n_\ell}}{\sqrt{n_\ell!}}
	\, | {\rm vac} \rangle \; ,
\label{eq:FOS}
\end{equation}
where $| {\rm vac} \rangle$ is the ``empty-lattice'' state. A possible
basis of the extended space would be provided by the products
$| \{ n_\ell \} \rangle \exp(i m\omega t)$ with integer $m$. However, it is
more useful here to employ the basis of Floquet-Fock states (or ``dressed
Fock states'') given by
\begin{equation}
	| \{ n_\ell \},m\rangle\!\rangle \equiv | \{ n_\ell \} \rangle
	\exp\!\left\{-\ri \frac{K}{\hbar\omega}\sin(\omega t)
	\sum_\ell \ell n_\ell + \ri m \omega t \right\} \, ,
\label{eq:FFS}
\end{equation}
which already diagonalize $\hat{H}(t) - \ri \hbar \partial_t$ for vanishing
interwell hopping strength $J = 0$ \citep{EckardtEtAl05b,EckardtHolthaus07}.
Invoking the scalar product~(\ref{eq:SCP}), in this basis the matrix of the
quasienergy operator has the structure
\begin{eqnarray}
	\langle \! \langle \{ n_\ell' \} , m' |
	\, \hat{H}(t) - \ri\hbar\partial_t \,
	| \{ n_\ell \} , m \rangle \! \rangle	
 	& = & \delta_{m',m} \, \langle \{ n_\ell' \} |
	\, \hat{H}_{\rm eff} + m\hbar\omega \, | \{ n_\ell \} \rangle
\nonumber \\	& &
	+ (1 - \delta_{m',m}) \,
	\langle \{ n_\ell' \} | \, \hat{V} \, | \{ n_\ell \} \rangle \; ,
\label{eq:QEM}
\end{eqnarray}
where $\hat{H}_{\rm eff}$ is a time-independent Bose-Hubbard Hamiltonian of
the familiar form~(\ref{eq:BHH}), but with the hopping matrix element~$J$
multiplied by a zero-order Bessel function ${\rm J}_0$ with the argument 
$K_0 = K/(\hbar\omega)$, resulting in the effective hopping strength
\begin{equation}
	J_{\rm eff} = J \, {\rm J}_0 \Big(K/(\hbar\omega)\Big) \; .
\label{eq:MHM}
\end{equation}	
The operator~$\hat V$, which is off-diagonal with respect to the photon
index~$m$, contains nearest-neighbor couplings~$J$ scaled by higher-order
Bessel functions $\pm {\rm J}_{m - m'}\!\big( K/(\hbar\omega) \big)$
\citep{EckardtEtAl05b,EckardtHolthaus07}. Thus, the diagonalization of the
quasienergy matrix~(\ref{eq:QEM}) constitutes a tremendous problem of a kind
not usually considered in traditional many-body physics: Infinitely many
$\hat{H}_{\rm eff}$-blocks, each one corresponding to the full Hamiltonian
matrix of a time-independent Bose-Hubbard model~(\ref{eq:BHH}) with modified
hopping matrix elements~(\ref{eq:MHM}), are shifted against each other
in energy by integer multiples of $\hbar\omega$, and are coupled by
$\hat{V}$-blocks. The latter blocks, in their turn, embody hopping elements
which are multiplied by Bessel-function factors with indices reflecting the
distance of the respective block from the main diagonal. In effect, these
$\hat{V}$-couplings cause multiphoton-like resonances among the states
described by the shifted diagonal blocks \citep{EckardtHolthaus08}. While
the physics of this problem has not yet been explored in full generality, its
high-frequency regime is comparatively transparent. Namely, if $\hbar\omega$
is large compared to the two energy scales $J$ and $U$ (while remaining still
small compared to the band gap~$\Delta$ of the underlying optical lattice, so
that the single-band treatment remains applicable), it is a good approximation
to neglect the couplings induced by $\hat V$ altogether, so that the driven
system~(\ref{eq:DBH}) reduces to an undriven system described by
$\hat{H}_{\rm eff}$, meaning that the effect of the time-periodic force
essentially is to ``renormalize'' the hopping matrix element according to
Eq.~(\ref{eq:MHM}). With the $\hat V$-couplings out of action, all
$\hat{H}_{\rm eff}$-blocks are equivalent, and returning from the extended
to the physical Hilbert space is tantamount to considering only one such 
block from the outset. Since the ratio $J/U$ governs the superfluid-to-Mott
insulator transition intrinsic to the Bose-Hubbard model
\citep{FisherEtAl89,Zwerger03,BlochEtAl08}, the renormalization of $J$ in 
response to high-frequency forcing implies that it is possible to induce that 
transition through adjusting the parameters of the driving force, while 
keeping the lattice depth and hence $U$ constant \citep{EckardtEtAl05b,
EckardtHolthaus08a}. This type of coherent control over the superfluid-to-Mott 
insulator transition has been demonstrated in a pioneering experiment by 
\citet{ZenesiniEtAl09}. It involves adiabatic following of the
$\hat{H}_{\rm eff}$-ground state when the driving amplitude is slowly changed; 
this has meanwhile been studied in detail with the help of numerical
simulations by \citet{PolettiKollath11}.

As in the time-independent case, experimental information about the many-body
state $|\psi(t)\rangle$ is obtained by time-of-flight absorption imaging:
Switching off the time-periodically shaken lattice potential at some 
moment~$t$, then letting the matter wave expand (neglecting interaction effects
during that expansion), and finally recording its density in space after a 
sufficiently long expansion time yields the momentum distribution 
\citep{Zwerger03,BlochEtAl08}
\begin{eqnarray}
	n(p,t) & = &
	\langle \psi(t) |
	\hat{a}^{\dagger}\!(p) \hat{a}(p)
	| \psi(t) \rangle	
\nonumber \\	& = &
	|\widetilde{w}(p)|^2 \sum_{r,s}
	\exp\left( \frac{\ri}{\hbar}(r - s)pd \right)
	\langle \psi(t) |
	\hat{b}^\dagger_r \hat{b}^{\phantom\dagger}_s
	| \psi(t) \rangle \, ,
\label{eq:MOD}
\end{eqnarray}	
where $\hat{a}(p)$ is the annihilation operator for a free-particle state
with momentum~$p$ in the direction of the lattice, and $\widetilde{w}(p)$ is
the Fourier transform of the Wannier function pertaining to the lowest Bloch
band; again, $d = \lambda/2$ denotes the lattice constant. Thus, apart from
the factor $|\widetilde{w}(p)|^2$ the observed momentum distribution is given 
by the Fourier transform of the system's one-particle density matrix $\langle 
\psi(t) | \hat{b}^\dagger_r \hat{b}^{\phantom\dagger}_s | \psi(t) \rangle$.
Let us now assume that for a given driving amplitude $K/(\hbar\omega)$, and 
hence for a given value of $J_{\rm eff}$, an eigenstate of $\hat{H}_{\rm eff}$ 
with energy~$\varepsilon$ takes the form
\begin{equation}
	| \psi_{\rm eff}(t) \rangle = \exp(-\ri \varepsilon t/\hbar)
	\sum_{\{n_\ell\}} \gamma_{\{n_\ell\}} | \{ n_\ell \} \rangle
\label{eq:EES}
\end{equation}	
with certain coefficients $\gamma_{\{n_\ell\}}$. Under conditions such that
the high-frequency approximation detailed above is valid, the corresponding
many-body state in the driven optical lattice is then obtained by replacing
the Fock states~(\ref{eq:FOS}) in this expansion~(\ref{eq:EES}) by the
Floquet-Fock states~(\ref{eq:FFS}) with $m = 0$, say, resulting in
\begin{equation}
	| \psi(t) \rangle = \exp(-\ri \varepsilon t/\hbar)
	\sum_{\{n_\ell\}} \gamma_{\{n_\ell\}} | \{ n_\ell \} \rangle
	\exp\!\left\{-\ri \frac{K}{\hbar\omega}\sin(\omega t)
	\sum_\ell \ell n_\ell \right\} \; .
\label{eq:FLO}
\end{equation}
In order to evaluate the momentum distribution~(\ref{eq:MOD}) for this
particular many-body Floquet state~(\ref{eq:FLO}) we then compute
\begin{eqnarray}
	\langle \psi(t) |
	\, \hat{b}^\dagger_r \hat{b}^{\phantom\dagger}_s \,
	| \psi(t) \rangle
	& = & 	 	
	\sum_{ \{n_\ell\},\{n_\ell'\} }
	\gamma^*_{\{n_\ell'\}} \gamma_{\{n_\ell\}}
	\langle \{n_\ell'\} |
	\, \hat{b}^\dagger_r \hat{b}^{\phantom\dagger}_s \,
	| \{n_\ell\} \rangle
\nonumber \\ & & \times	
	\exp\!\left\{-\ri \frac{K}{\hbar\omega}\sin(\omega t)
	\sum_\ell \ell (n_\ell - n_\ell') \right\} \; .
\label{eq:SPC}
\end{eqnarray}	
Now the only nonvanishing matrix elements are those with
$n_\ell = n_\ell'$ for $\ell \ne r,s$; together with $n_r' = n_r + 1$
and $n'_s = n_s - 1$. This implies
\begin{eqnarray}
	\sum_\ell \ell(n_\ell - n_\ell') = -(r-s) \; ,
\end{eqnarray}
which allows us to take the phase factor appearing on the right-hand side
of Eq.~(\ref{eq:SPC}) out of the sum, giving
\begin{equation}
	\langle \psi(t) |
	\, \hat{b}^\dagger_r \hat{b}^{\phantom\dagger}_s \,
	| \psi(t) \rangle = 	
	\langle \psi_{\rm eff} |
	\, \hat{b}^\dagger_r \hat{b}^{\phantom\dagger}_s \,
	| \psi_{\rm eff} \rangle
	\exp\!\left\{\ri(r-s)\frac{K}{\hbar\omega}\sin(\omega t)
	\right\} \; .
\end{equation}	
Inserting this result into the representation~(\ref{eq:MOD}), we immediately
obtain the momentum distribution provided by a single Floquet state of the
driven Bose-Hubbard model~(\ref{eq:DBH}) in the high-frequency regime:
\begin{eqnarray}		
	n(p,t) & = &
	|\widetilde{w}(p)|^2 \sum_{r,s}
	\langle \psi_{\rm eff} |
	\, \hat{b}^\dagger_r \hat{b}^{\phantom\dagger}_s \,
	| \psi_{\rm eff} \rangle
\nonumber \\	& \times &		
	\exp\!\left\{ \frac{\ri}{\hbar}(r - s)
	\left( pd + \frac{K}{\omega} \sin(\omega t) \right) \right\} \; ,
\label{eq:ICF}
\end{eqnarray}	
where, by assumption, $|\psi_{\rm eff}\rangle$ is the associated energy
eigenstate of $\hat{H}_{\rm eff}$.

However, an important step is still missing in order to connect theory with
experiment: The driven Bose-Hubbard model, with the inertial force being
incorporated through the driving term~(\ref{eq:DCF}), refers to the co-moving
frame of reference as considered in Sec.~\ref{sec:S_3}, whereas measurements
usually are performed in the laboratory frame. Thus, in order to obtain 
the momentum distribution $n_{\rm lab}(p_{\rm lab},t)$ as recorded by an 
observer in the laboratory frame, we still have to invert the 
transformation~(\ref{eq:TRP}). This is done in the general case by writing
\begin{equation}
	p = p_{\rm lab} - M \dot{X}_0(t) \; ,
\label{eq:PLG}
\end{equation}
and through
\begin{equation}
	p = p_{\rm lab} - \frac{K}{\omega d} \sin(\omega t)
\label{eq:PIN}	
\end{equation}
for the driven Bose-Hubbard model with forcing~(\ref{eq:DCF}),
as corresponding to a frequency variation
$\Delta \nu(t) = \Delta \nu_{\rm max} \sin(\omega t)$
between the counterpropagating laser beams, or to lattice motion
$\dot{X}_0(t) = L\omega\sin(\omega t)$,
keeping in mind the relation~(\ref{eq:MAF}) for the driving amplitude,
together with the definition $K = F_1 d$. Therefore, we finally have
\begin{eqnarray}
	n_{\rm lab}(p_{\rm lab},t) & = &
	\Big| \widetilde{w}\Big(p_{\rm lab} - K/(\omega d)\sin(\omega t)\Big)
	\Big|^2
\nonumber \\	& & \times	
	\sum_{r,s}\exp\left( \frac{\ri}{\hbar}(r - s)p_{\rm lab}d \right)
	\langle \psi_{\rm eff} |
	\, \hat{b}^\dagger_r \hat{b}^{\phantom\dagger}_s \,
	| \psi_{\rm eff} \rangle \; .
\label{eq:PAT}
\end{eqnarray}			
This is quite a significant observation: The momentum distribution of a matter
wave occupying a many-body Floquet state in a time-periodically shaken optical
lattice~(\ref{eq:OSL}) in the high-frequency regime equals that of the
associated energy eigenstate of $\hat{H}_{\rm eff}$, which is obtained from
the undriven Bose-Hubbard Hamiltonian~(\ref{eq:BHH}) through replacing the
hopping matrix element~$J$ by $J_{\rm eff}$ as defined in Eq.~(\ref{eq:MHM}),
insofar as exactly the same Fourier transform of the one-particle density 
matrix appears in both cases. Hence, even though the position of the lattice 
is periodically shifted, the peak pattern observed in the laboratory frame 
does not move. The effect of the time-periodic shift is seen only in the 
{\em envelope\/} of that pattern, given by the Fourier transform $\widetilde{w}$
of the Wannier function, the argument of which is modulated periodically in
time in accordance with Eq.~(\ref{eq:PIN}). Thus, apart from this modulation
the experimental signature of the superfluid-to-Mott insulator transition
occurring in shaken optical lattices upon changing the driving
amplitude \citep{ZenesiniEtAl09} is the same as that of the transition
occuring in a stationary lattice in response to a variation of its
depth \citep{Zwerger03,BlochEtAl08}.

It may be useful to point out that the experimental signatures differ from
the above description if the optical lattice is not shaken, but kept at rest,
while the force is induced by means of the time-periodic modulation of a
levitation gradient which ``stirs'' the condensate, as done in the experiments
by \citet{HallerEtAl10}. In this latter situation the model~(\ref{eq:DCF})
actually describes the driving force in the very reference frame in which
the momentum distribution is recorded, so that one obtains an oscillating
interference pattern given directly by Eq.~(\ref{eq:ICF}). In contrast, 
in the experiments reported by \citet{LignierEtAl07}, \citet{SiasEtAl08}, 
\citet{EckardtEtAl09}, \citet{ZenesiniEtAl09}, and in Sec.~\ref{sec:S_6} below,
the lattice is shaken in the laboratory frame according to Eq.~(\ref{eq:OSL}). 
Hence, here the ``micromotion'' is taken out of the interference pattern by 
means of Eq.~(\ref{eq:PIN}), which connects the momentum~$p$ in the co-moving 
frame to the momentum $p_{\rm lab}$ observed in the laboratory frame. The fact 
that the resulting interference pattern does not move in the laboratory frame 
\citep{EckardtEtAl05b,EckardtHolthaus07,EckardtHolthaus08a} may sometimes 
facilitate its interpretation.

\subsection{Signatures of ideal shaken Bose gases}

With respect to the Floquet condensates envisioned in Sec.~\ref{sec:S_2},
the case of an ideal Bose-Einstein condensate in a shaken optical lattice
now is of particular interest. For vanishing interaction $U = 0$ the
time-dependent Schr\"odinger equation for the model~(\ref{eq:DBH}) can easily
be solved exactly for any type of forcing described by a driving term
\begin{equation}	
	\hat{H}_1(t) = -F(t) d \sum_\ell \ell \hat{n}_\ell \; ,
\end{equation}
without requiring a specific time-dependence of the force. Therefore, we
abandon time-periodic forces and the Floquet picture for the moment being, and
consider an initial-value problem instead: We assume that $F(t)$ vanishes for
$t < 0$, and is switched on sharply at $t = 0$, but is arbitrary otherwise.
Generalizing the previous Eq.~(\ref{eq:MOF}), we thus impose a force
\begin{equation}
	F(t) = -M\ddot{X}_0(t) \, \Theta(t) - M \dot{X}_0(0+) \, \delta(t) \; ,
\label{eq:GDF}
\end{equation}
where the second term accounts for the sudden velocity jump of the lattice
from $\dot{X}_0(0-) = 0$ to an arbitrary value $\dot{X}_0(0+)$ at $t = 0$.
Moreover, we consider a lattice with $M_L \gg 1$ sites and disregard
finite-size effects, so that the operator
\begin{equation}
	\hat{c}^\dagger_k(0) = \frac{1}{\sqrt{M_L}} \sum_\ell
	\exp(\ri\ell kd) \, \hat{b}^\dagger_\ell
\end{equation}
creates a particle in the Bloch state with quasimomentum $\hbar k$. We then
assume that the initial state at $t = 0$ is an ideal $N$-particle condensate
occupying such a Bloch state. Even if $k = 0$ might be the only experimentally
realistic option here, we do not impose this restriction at this point. For
$t > 0$, after the force has been turned on, the resulting $N$-particle wave
function can then be written in the form
\begin{equation}
	| \psi_{k}(t) \rangle = \frac{1}{\sqrt{N!}}
	\left[ \hat{c}^\dagger_{k}(t) \right]^N
	| {\rm vac} \rangle \; .
\label{eq:NPH}
\end{equation}
Here the creation operator $\hat{c}^\dagger_{k}(t)$, given by
\begin{equation}
	\hat{c}^\dagger_{k}(t) = \exp\!\left(-\frac{\ri}{\hbar}
	\int_0^t \! \rd \tau \, E\Big(q_{k}(\tau)\Big) \right)
	\frac{1}{\sqrt{M_L}} \sum_\ell
	\exp\!\big(\ri \ell q_{k}(t) d\big) \,
	\hat{b}^\dagger_\ell \; ,
\label{eq:CRH}
\end{equation}	
refers to a so-called Houston state, also known as accelerated Bloch state
\citep{Houston40,EckardtEtAl09}. This nomenclature stems from the fact that
the time-dependent wave number $q_{k}(t)$ appearing here has to obey the
``semiclassical'' acceleration law
\begin{equation}
	\hbar \dot{q}_{k}(t) = F(t) \;
\label{eq:ACT}	.	
\end{equation}
Therefore, using the particular connection~(\ref{eq:GDF}) between the inertial
force and the lattice motion, and requiring that $q_{k}(t)$ be equal to the
wave number~$k$ of the initial state for $t < 0$, we have
\begin{eqnarray}
	q_{k}(t) & = &
	k + \frac{1}{\hbar}\int_{0}^t \! \rd \tau \, F(\tau)
\nonumber\\    & = &
	k - \frac{M}{\hbar}\Big( \dot{X}_0(t) - \dot{X}_0(0+) \Big)
	- \frac{M}{\hbar} \dot{X}_0(0+)
\nonumber\\	& = &
	k - \frac{M}{\hbar} \dot{X}_0(t) 	 		
\label{eq:QKT}
\end{eqnarray}	
for $t > 0$, having properly accounted for the delta-kick at the moment of
turn-on. Finally, the expression   	
\begin{equation}
	E(k) = -2J \cos(kd)
\label{eq:EDR}	
\end{equation}
appearing in the exponential of Eq.~(\ref{eq:CRH}) denotes the single-particle
dispersion relation describing the energy band provided by the
Hamiltonian~(\ref{eq:BHH}) when $U = 0$.

For such a noninteracting $N$-particle wave function~(\ref{eq:NPH}) the
momentum distribution~(\ref{eq:MOD}) is given, apart from the factor
$|\widetilde{w}(p)|^2$, by
\begin{eqnarray}
	& &
	\sum_{r,s} \exp\!\Big(\ri(r - s)pd/\hbar\Big) \,
	\langle \psi_{k}(t) |
	\hat{b}^\dagger_r \hat{b}^{\phantom\dagger}_s
	| \psi_{k}(t) \rangle
\nonumber \\ & = &
	N \sum_r \exp\Big(\ri r[p/\hbar - q_{k}(t)]d\Big) \; .
\label{eq:IFP}
\end{eqnarray}	
The return to the laboratory frame now is achieved with the help of
Eq.~(\ref{eq:PLG}), and the expression~(\ref{eq:QKT}) for $q_{k}(t)$:
Using these, one obtains
\begin{equation}
	p/\hbar - q_{k}(t) = p_{\rm lab}/\hbar - k  \; .
\label{eq:PLB}	
\end{equation}
Therefore, the interference pattern provided by an ideal Houston condensate
in time-of-flight absorption imaging again is stationary in the laboratory
frame, and the peak positions are not affected by the force: Even after the
force~(\ref{eq:GDF}) has been switched on, the interference peaks are still
permanently centered around the wave numbers $k \bmod (2\pi/d)$, as they had
been for the unforced initial condensate. It needs to be stressed that this
peculiar feature is crucially dependent on the delta-kick which accompanies
the sudden turn-on of the force in the co-moving frame.

In particular, let us now consider a monochromatic oscillating force switched
on instantaneously at $t = 0$ with starting phase $\phi$ according to
Eq.~(\ref{eq:MOF}), which leads to
\begin{equation}
	q_{k}(t) =
	k - \frac{F_1}{\hbar\omega}\sin(\omega t + \phi)
\label{eq:QPT}
\end{equation}
for $t > 0$. In this case there is a close relation between the $N$-particle
Houston states~(\ref{eq:NPH}), which have been constructed for $t > 0$ as
solutions of an initial-value problem, and $N$-particle Floquet states, which
presuppose a perfectly time-periodic force $F(t) = -F_1\cos(\omega t + \phi)$
acting at {\em all\/} times~$t$. In order to obtain these Floquet states, one
only has to extend $q_{k}(t)$, as given by Eq.~(\ref{eq:QPT}), to all~$t$,
and use this expression in Eq.~(\ref{eq:CRH}), now to be considered for
all~$t$. Then the emerging ``extended'' Houston states~(\ref{eq:NPH}) have
the basic form~(\ref{eq:FLS}): By construction, $q_{k}(t)$ is $T$-periodic,
with $T = 2\pi/\omega$, but the energy integral in the exponent of
Eq.~(\ref{eq:CRH}) is not. However, writing
\begin{eqnarray*}
	& &
	\exp\!\left(-\frac{\ri}{\hbar} \int_0^t \! \rd \tau \,
	E(q_{k}(\tau)) \right)
\nonumber \\ & = &
	\exp\!\left(-\frac{\ri}{\hbar} \int_0^t \! \rd \tau \,
	\big[ E(q_{k}(\tau)) - \varepsilon(k) \big] \right)
	\exp\Big(-\ri\varepsilon(k) t/\hbar\Big)
\nonumber
\end{eqnarray*}
with the help of the one-cycle-averaged energies
\begin{eqnarray}
	\varepsilon(k) & = &
	\frac{1}{T} \int_0^T \! \rd \tau \, E(q_{k}(\tau))
\nonumber \\	& = &
	- 2 J \, {\rm J}_0 \! \left(\frac{F_1d}{\hbar\omega}\right)
	\cos(k d) \; ,   	
\label{eq:WRL}
\end{eqnarray}		
the first of these exponentials is $T$-periodic and therefore part of the
Floquet functions, whereas the occurrence of the averages~(\ref{eq:WRL}) in
the second exponential, accompanied by the time~$t$, allows one to identify
them as quasienergies. The Houston-Floquet states thus found are particularly
simple examples of spatiotemporal Bloch waves, incorporating both the spatial
periodicity of the lattice and the temporal periodicity of the driving force
on equal footing. They are labeled by the same quantum numbers $k$ as the 
customary, time-independent Bloch waves to which they reduce in the absence
of the drive, while their time-evolution, apart from the time-periodic motion 
incorporated into the moving wave numbers $q_k(t)$, is specified by the 
quasienergies $\varepsilon(k)$ \citep{ArlinghausHolthaus11}. Observing that 
one again encounters here the effective hopping matrix element $J_{\rm eff}$ 
introduced in Eq.~(\ref{eq:MHM}), the quasienergy-quasimomentum dispersion 
relation takes the form
\begin{equation}
	\varepsilon(k) = - 2 J_{\rm eff} \cos(kd)
	\quad \bmod \hbar \omega \; ,
\label{eq:QED}
\end{equation}	
which differs from the original dispersion relation~(\ref{eq:EDR}) of the
undriven lattice only through the replacement of $J$ by $J_{\rm eff}$.
That same replacement had been met before in the context of the interacting
many-body system described by the driven Bose-Hubbard model~(\ref{eq:DBH}),
when constructing the approximate effective Hamiltonian~$\hat{H}_{\rm eff}$
pertaining to the high-frequency regime. In contrast, in the noninteracting
case considered here no approximations have been made; Eq.~(\ref{eq:QED})
holds exactly for all driving frequencies.

The fact that the quasienergies~(\ref{eq:QED}) for the driven,
noninteracting Bose-Hubbard model could be calculated by taking the
time-averages~(\ref{eq:WRL}) further suggests that these quasienergies do
coincide with the corresponding mean energies~(\ref{eq:FME}), as considered
in Sec.~\ref{sec:S_2}. Indeed, taking a single-particle Houston-Floquet state
$|\psi_k(t)\rangle = |u_k(t)\rangle\exp\Big(-\ri\varepsilon(k)t/\hbar\Big)$
as constructed above, one easily confirms the identity
\begin{equation}
	\overline{E}(k)
	= \langle \! \langle u_k | \hat{H}(t) | u_k \rangle \! \rangle
	= - 2 J_{\rm eff} \cos(kd) \; ,
\label{eq:HFM}
\end{equation}
so that here the mean energies of the Floquet states actually equal the
quasi\-energies~(\ref{eq:QED}), disregarding their
``$\bmod \, \hbar \omega$''-multiplicity. In the case of a general $T$-periodic
single-particle Hamiltonian $H^{(1)}(t)$ with Floquet states~(\ref{eq:FLS})
this is not the case, since then
\begin{eqnarray}
	\overline{E}_n & = & \langle \! \langle u_n |
	H^{(1)} - \ri\hbar\partial_t | u_n \rangle \! \rangle +
	\langle \! \langle u_n | \ri\hbar\partial_t | u_n \rangle \! \rangle
\nonumber \\	& = &
	\varepsilon_n +
	\langle \! \langle u_n | \ri\hbar\partial_t | u_n \rangle \! \rangle	
	\; ,	
\end{eqnarray}
as discussed in detail by \citet{FainshteinEtAl78}.

Interestingly, the quasienergy band~(\ref{eq:QED}) ``collapses'' when
$K_0 = F_1 d/(\hbar\omega)$ equals a zero of the ${\rm J}_0$ Bessel function
\citep{Holthaus92}, implying that an arbitrary single-particle wave packet
driven under such conditions cannot spread, but reproduces itself
$T$-periodically. This phenomenon, termed ``dynamic localization''
\citep{DunlapKenkre86}, has recently been observed with dilute Bose-Einstein
condensates in driven optical lattices
\citep{LignierEtAl07,EckardtEtAl09,ArlinghausEtAl11}. With a view towards
future applications, the dependence of the quasienergy band width on the
driving strength also has been identified as a means of controlling transport
in systems with attractive pairing interactions \citep{KudoEtAl09}.

Once again, it is instructive to compare the above results, obtained for a
shaken optical lattice, to the corresponding physics when the lattice is at
rest, while the condensate is stirred by a harmonically modulated levitation
gradient \citep{HallerEtAl10}. Then it is actually possible to impose an
instantaneously turned-on oscillating force
\begin{equation}
	\widetilde{F}(t) = - F_1 \cos(\omega t + \phi) \, \Theta(t)
\label{eq:STF}
\end{equation}
without the additional delta-kick present in Eq.~(\ref{eq:MOF}), which
necessarily appears when an inertial force is abruptly turned on. The 
moving wave numbers $q_k(t)$, which previously had been determined in 
Eq.~(\ref{eq:QKT}) as solutions to the equation of motion~(\ref{eq:ACT}), 
now have to be replaced by the solutions $\widetilde{q}(t)$ to the 
corresponding equation $\hbar\dot{\widetilde{q}}(t) = \widetilde{F}(t)$, 
giving
\begin{eqnarray}
	\widetilde{q}(t) & = &
	k + \frac{1}{\hbar}\int_{0}^t \! \rd \tau \, \widetilde{F}(\tau)
\nonumber\\    & = &
	k + \frac{F_1}{\hbar\omega}\sin(\phi)
	- \frac{F_1}{\hbar\omega}\sin(\omega t + \phi) 	
\end{eqnarray}
for $t > 0$, instead of Eq.~(\ref{eq:QPT}) above. Therefore, one formally has
\begin{equation}
	\widetilde{q}(t) = q_{k + \Delta k}(t)
\end{equation}
with a wave-number shift
\begin{equation}
	\Delta k = \frac{F_1}{\hbar\omega}\sin(\phi) \; ,
\end{equation}	
which means that an additional momentum $\hbar\Delta k$ is imparted on the
condensate particles when the stirring force is switched on rapidly, as
described by Eq.~(\ref{eq:STF}). Thus, besides the micromotion the resulting
time-of-flight absorption images would also show a shift of the peak positions.
Both of these features are absent in experiments with shaken lattices: It
really matters whether the condensate is ``shaken'' or ``stirred''!    	

We note that \citet{KudoMonteiro11a,KudoMonteiro11b}, focusing on stirring
forces of the form~(\ref{eq:STF}), have suggested to introduce effective
dispersion relations which also incorporate the initial phase $\phi$. While
one is free to adopt this viewpoint, it seems to obscure the conceptual 
simplicity of the Floquet approach: The quasienergy band~(\ref{eq:QED}) 
characterizes the time-periodically driven system as such, regardless of the 
way the drive has been switched on, that is, independent of the phase~$\phi$ 
which parametrizes the sudden turn-on~(\ref{eq:STF}), or of any other 
parameters which specify other, equally possible turn-on protocols. Because 
this quasienergy band~(\ref{eq:QED}) actually consists of eigenvalues of the 
quasienergy operator it even provides a dispersion relation in the usual
sense of solid-state physics which allows one to compute group velocities by
taking its derivative, properly evaluated at that quantum number~$k$ around
which the wave packet is centered \citep{ArlinghausHolthaus11}.

With regard to shaken condensates, we may summarize our considerations
as follows: If a perfectly ideal Bose-Einstein condensate in an optical
lattice initially occupies a Bloch state with wave number~$k$, and then is
abruptly being shaken and thus subjected to the force~(\ref{eq:MOF}) under
single-band conditions, it permanently populates a single Floquet state
labeled by the same wave number $k$, regardless of the phase $\phi$.

\section{Experimental findings}
\label{sec:S_6}

\begin{figure}[t]
\centerline{\includegraphics[angle=0., scale = 0.7, width = 0.9\linewidth]
	{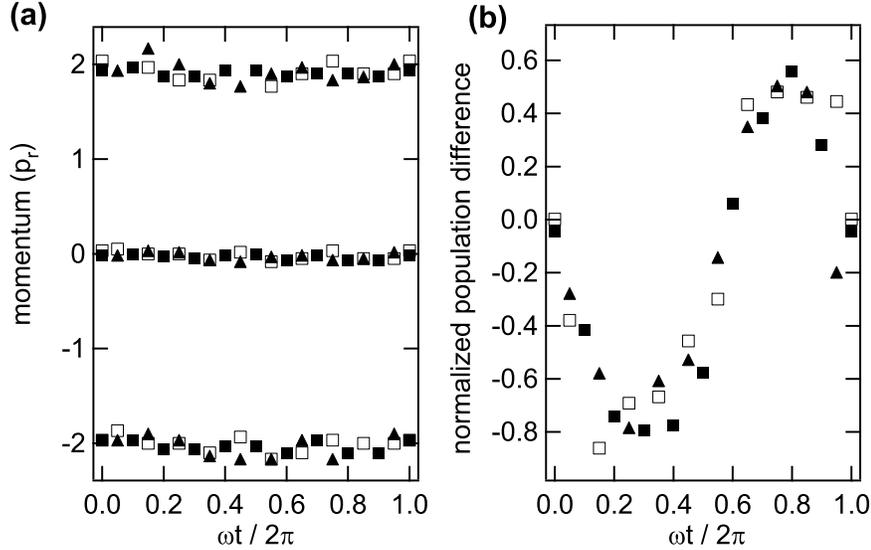}}
\caption[Fig.~2]{(a) Position of interference maxima observed in time-of-flight
 	absorption imaging of condensates released from an optical lattice with
	depth $V_0 = 9 \, \Er$, shaken with frequency $\omega/2\pi = 1$~kHz and
	scaled amplitude $K/(\hbar\omega) = 1.85$, during one shaking cycle. 
	The starting phases $\phi$ of the drive are $0$ (solid squares), 
	$+\pi/2$ (open squares), and $-\pi/2$ (solid triangles); the images 
	were taken 10 driving cycles after its turn-on. The maxima are 
	centered around positions corresponding to momenta 
	$p_{\rm lab} = 0, \pm2 \, p_{\rm r}$.
	(b) Modulation of the population difference of the side peaks during 
	one shaking cycle. All three starting phases (symbols as in (a)) give 
	rise to the same curve.}
\label{F_2}
\end{figure}

In order to substantiate the relevance of the above ideal-gas considerations 
for laboratory experiments, we took time-of-flight absorption images of dilute 
$^{87}$Rb-condensates released from shaken 1D optical lattices~(\ref{eq:OSL}), 
employing various driving frequencies and amplitudes. In Fig.~\ref{F_2} we 
display typical results obtained for a lattice with depth $V_0 = 9 \, \Er$, 
driven with frequency $\omega/2\pi = 1$~kHz and scaled amplitude 
$K/(\hbar\omega) = 1.85$. Under such conditions the single-band approximation 
is well justified; according to Eq.~(\ref{eq:MHM}), one has 
$J_{\rm eff} = 0.311 \, J$. The drive was turned on quickly with a starting 
phase $\phi$, as modeled by Eq.~(\ref{eq:MOF}), and the images were taken 
10~driving cycles later. Hence, provided that the solution~(\ref{eq:NPH}) to 
the initial-value problem for the ideal gas discussed in Sec.~\ref{sec:S_5} 
describes the real laboratory system correctly, Eqs.~(\ref{eq:IFP}) and 
(\ref{eq:PLB}) tell us that for {\em any\/} value of $\phi$ there should be 
stationary interference maxima at $p_{\rm lab} = 0 \bmod \hbar(2\pi/d)$, or
\begin{equation}
	\frac{p_{\rm lab}}{p_{\rm r}} = 0 \bmod 2 \; ,
\end{equation}	
using the recoil momentum $p_{\rm r} = \hbar\kL$ as reference scale. This
expectation is fully confirmed in Fig.~\ref{F_2}. In all three cases
considered there, with $\phi = 0$, $+\pi/2$, and $-\pi/2$, the central peak
is located at $p_{\rm lab} = 0$, as corresponding to a condensate occupying
the Houston-Floquet state associated with the {\em minimum\/} of the
quasienergy band~(\ref{eq:QED}). In addition there are side peaks at
$p_{\rm lab} = \pm 2 \, p_{\rm r}$, corresponding to the
width $\hbar (2\pi/d) = 2 \, p_{\rm r}$ of the quasimomentum Brillouin zone.
The left panel of Fig.~\ref{F_2} depicts the time-resolved evolution of the
positions of the respective interference maxima during a single cycle of the
driving force. As anticipated, these positions remain practically constant
in time, apart from apparent slight wigglings. The right panel of 
Fig.~\ref{F_2} shows the modulation of the height of the side peaks during
one driving cycle, which stems from the oscillating time-dependence of the
argument $p$ of the factor $|\widetilde{w}(p)|^2$ determining the envelope
of the interference pattern in accordance with Eq.~(\ref{eq:PAT}). For all 
values of the starting phase we observe practically the same signals, which 
is in line with the surmise that they are signatures of the same state.

Therefore, we interpret the data displayed in Fig.~\ref{F_2} as experimental
signatures of Floquet condensates: The existence of a stationary, stable and
lasting interference pattern indicates that the driven Bose gas actually
tends to occupy a single Houston-Floquet state. The fact that this pattern
is centered around $p_{\rm lab} = 0 \bmod 2 \, p_{\rm r}$ implies that this
state is the one at the bottom of the quasienergy band~(\ref{eq:QED}), which,
by virtue of Eq.~(\ref{eq:HFM}), also is the one equipped with the lowest
mean energy~(\ref{eq:FME}). This is not surprising, because we are starting 
with a condensate occupying the Bloch state $k = 0$, and because a drive of 
the form~(\ref{eq:MOF}) does not change the quantum number~$k$ when it is 
turned on.

\begin{figure}[t]
\centerline{\includegraphics[angle=0., scale = 0.5, width = 0.8\linewidth]
	{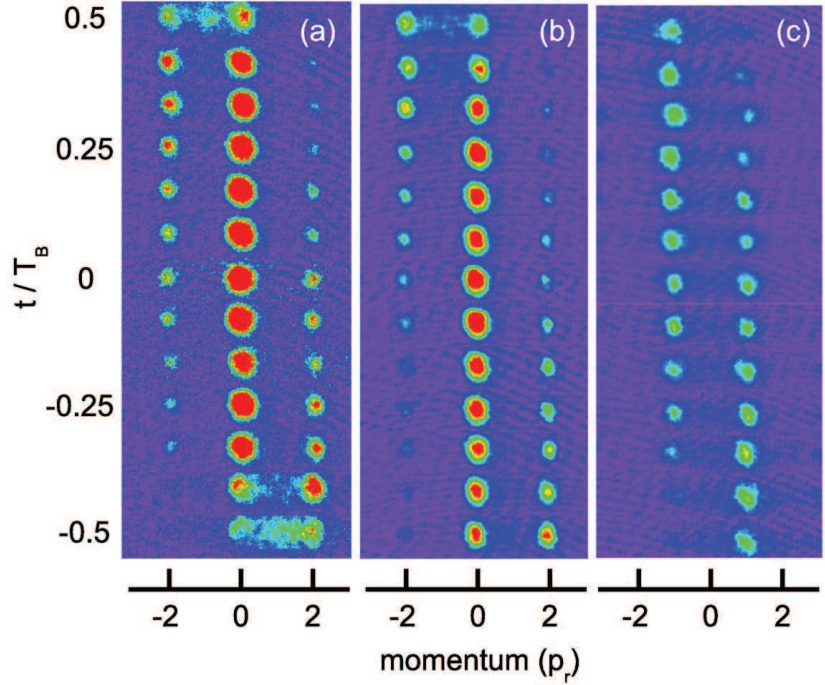}}
\caption[Fig.~3]{(Color online) Bloch oscillations in driven optical lattices
	($V_0/\Er = 10$), recognizable as shifts of the interference pattern
	in time by one Brillouin zone of width $2 \, p_{\rm r}$ within 
	one Bloch cycle $T_{\rm B}$. The Bloch frequency is set to 
	$\omega_{\rm B}/2\pi = 242.4$~Hz, whereas the driving frequency is 
	$\omega/2\pi = 3$~kHz. The dimensionless driving amplitudes 
	$K/(\hbar\omega)$ correspond to 0, 1.5, and 4 (left to right). 
	In (a), and to a lesser extent in (b), the pattern appears slightly 
	smeared at times $t/T_{\rm B} = \pm 1/2$, due to the occurrence of 
	a dynamical instability. Note that the peak positions in (c) are 
	centered around the edges $\pm p_{\rm r} = \pm\hbar\pi/d$ of the 
	quasimomentum Brillouin zone.}
\label{F_3}
\end{figure}

If a Floquet condensate is to be regarded as an entity of its own, then it
should behave as such when probed by a weak force acting on top of the
time-periodic, driving one. This is actually the case, as illustrated by a
sequence of further measurements summarized in Fig~\ref{F_3}: Here we consider
a lattice with depth $V_0 = 10 \, \Er$ driven by an oscillating force with an
additional weak static component, written as
\begin{equation}
	F(t) = \Big( -F_0 - F_1 \cos(\omega t) \Big) \Theta(t) \; ,
\label{eq:FSO}	
\end{equation}
so that the time-dependent wave numbers~(\ref{eq:QKT}) now become
\begin{equation}
	q_k(t) = k - \frac{F_0t}{\hbar}
	- \frac{F_1}{\hbar\omega}\sin(\omega t)
\end{equation}
for $t > 0$, disregarding an initial phase $\phi$ from the outset. Therefore,
the system undergoes a pure Bloch oscillation when $F_1 = 0$: In response to
a static force of strength $F_0$, one wave-number Brillouin zone of width
$2\pi/d$ is then traversed at uniform ``speed'' $F_0/\hbar$ within the Bloch
time
\begin{equation}
	T_{\rm B} = \frac{2\pi\hbar}{F_0d} \; ,
\end{equation}
and the periodicity of the energy-quasimomentum relation~(\ref{eq:EDR}) in
$k$ gives rise to an oscillating wave-packet motion, as long as interband
transitions remain negligible \citep{Zener34}. We fix the Bloch frequency
$\omega_{\rm B}/2\pi = 1/T_{\rm B}$ at $242.4$~Hz, and the driving frequency
$\omega/2\pi$ at $3.0$~kHz, more than 12 times higher. This clear separation
of time scales implies transparent dynamics: Essentially, the oscillating
component of the force~(\ref{eq:FSO}) ``dresses'' the gas, making it
condense into the Floquet state associated with the minimum of the
quasienergy-quasimomentum relation~(\ref{eq:QED}), as before. The relatively
weak static component then merely probes this dressed system, making it behave
as an undressed would if Eq.~(\ref{eq:QED}) actually were the {\em energy\/}
dispersion relation \citep{ArlinghausHolthaus11}.

The signatures of such dynamics are again visible in time-of-flight
absorption images. We expect stationary interference peaks, corresponding
to the minima of the dispersion relation~(\ref{eq:QED}), while the argument
of the envelope-giving function $|\widetilde{w}(p)|^2$ in Eq.~(\ref{eq:PAT})
evolves in time as
\begin{equation}
	\frac{p}{p_{\rm r}} =
	\frac{p_{\rm lab}}{p_{\rm r}} - 2\frac{t}{T_{\rm B}}
	- \frac{K/(\hbar\omega)}{\pi} \sin(\omega t) \; ,
\end{equation}	
with $K = F_1 d$. This is precisely what is seen in Fig.~\ref{F_3}: Here
we show absorption images obtained from condensates released at times
$t_i/T_{\rm B} = i/12$ with $i = -6, -5, \ldots , 5, 6$, spanning one full 
Bloch cycle~$T_{\rm B}$ (negative times here formally correspond to negative 
forces $F_0$ in Eq.~(\ref{eq:FSO})). 
In the leftmost panel we consider $K/(\hbar\omega) = 0$, 
so that we are dealing with the bare system undergoing an undisturbed Bloch 
oscillation \citep{MorschEtAl01}, corresponding to an apparent displacement 
of the entire pattern by one Brillouin zone after one Bloch cycle. At 
$t/T_{\rm B} = \pm 1/2$, when $q_k(t) = \pm \pi/d$ and the system passes the 
zone boundaries, a familiar dynamical instability occurs \citep{ZhengEtAl04}, 
which becomes visible as a slight blurring of the peaks.

The middle panel shows the corresponding pictures
obtained for $K/(\hbar\omega) = 1.5$, so that $J_{\rm eff} = 0.512 \, J$:
Indeed this set of images looks quite similar to the previous one, validating
our view of a ``dressed condensate'' emerging through the application of the
strong oscillating component of the force~(\ref{eq:FSO}), and then responding
to the static component like a bare one with a renormalized hopping
matrix element~$J_{\rm eff}$. The effective reduction of $J$ by a factor of
about one half also leads to a notable reduction of the dynamical instability
\citep{ZhengEtAl04}.

The third panel finally shows the images obtained for $K/(\hbar\omega) = 4$,
resulting in a {\em negative\/} effective hopping element,
$J_{\rm eff} = -0.397 \, J$. This means that the minima of the quasienergy
dispersion~(\ref{eq:QED}) are now centered at the zone boundaries
$k = \pm \pi/d$, where the maxima of the original dispersion~(\ref{eq:EDR})
had been. Somewhat surprisingly, the gas spontaneously condenses into the
associated Floquet state, leaving its signature in the form of two interference
maxima neatly centered around $p = \pm p_{\rm r}$. This strongly dressed system
then again performs a Bloch oscillation, as wittnessed by the already familiar
apparent displacement of the entire pattern by one quasimomentum Brillouin zone
per Bloch cycle.

Thus, our system actively seems to select the respective Floquet state with 
the lowest mean energy, and seems to condense ``by itself'' into that state,
as has also been reported by \citet{LignierEtAl07} for the case of purely 
sinusoidal driving. This finding goes beyond the ideal-gas picture, and 
requires further studies. It is underlined by a related observation: If we 
instantaneously change the driving amplitude such that the absolute magnitude 
of $J_{\rm eff}$ is preserved, but its sign is reversed, the original
interference pattern caused by the initial state gradually vanishes, and the 
new pattern signaling the target state at the opposite band edge establishes 
itself within a few driving cycles. This process occurs in either direction, 
and appears to be fully reversible. This finding also implies that the 
dressed condensates settling down at the boundaries of the Brillouin zone 
when $J_{\rm eff}$ is negative actually are {\em stable\/}.

Such experiments involving a sudden change of $J_{\rm eff}$ are reminiscent
of ``quenching'' experiments in which a parameter of a time-independent
trapping potential is varied instantaneously, and the ensuing relaxation
dynamics are observed \citep{Dziarmaga10,PolkovnikovEtAl11}. However, it is
uncertain at this point whether our observations fit into this picture, and 
the underlying relaxation mechanism --- if it really is one --- needs to be 
understood in detail, keeping in mind the fact that the system is isolated 
and thus cannot get rid of excess energy. Still, in view of the preliminary 
thoughts put forward in Sec.~\ref{sec:S_2}, an interesting possibility 
suggests itself: The equilibrium state approached by a time-periodically 
driven, isolated Bose gas is determined not by usual thermodynamics, but by 
``periodic thermodynamics'' in the sense of \citet{Kohn01}, possibly involving 
a generalized temperature~$\Theta$ which depends on --- and can, therefore, 
be controlled by --- the external drive.

\section{Conclusions}
\label{sec:S_7}

In a nutshell, the main results of the present study are encoded in
Fig.~\ref{F_3}: The observation that the middle panel of this figure closely
resembles the leftmost one indicates that there are Floquet condensates,
that is, macroscopically occupied Floquet states of time-periodically forced
Bose gases; the observation that the interference maxima are centered at the
Brillouin zone edges when the effective hopping matrix element is negative,
as seen in the rightmost panel, indicates that the Floquet condensate is
carried by the Floquet state with the lowest mean energy. It remains to be
seen whether these conclusions, drawn from one particular laboratory setting,
also hold under more general circumstances.

The observation of stable interference patterns in the presence of the
time-periodic forcing gives strong support to the surmise, formulated in
Sec.~\ref{sec:S_2}, that time-periodically driven, weakly interacting Bose
gases confined by trapping potentals which forbid their escape effectively
constitute {\em equilibrium\/} systems, rather than nonequilibrium ones;
the equilibrium state being characterized by a constant distribution of
Floquet-state occupation numbers. Under conditions such that the mean-energy
constraint~(\ref{eq:MEC}) is valid, the expected distribution is the 
Bose-Einstein distribution~(\ref{eq:BED}), with the mean energies of the 
single-particle Floquet states replacing the single-particle energies appearing in
time-independent situations. That distribution is characterized by two
parameters $\alpha$ and $\gamma$, corresponding to the generalized chemical
potential $\nu$ and to the generalized temperature $\Theta$ introduced in 
Eqs.~(\ref{eq:DTH}) and (\ref{eq:DNU}). If this suggestion could be confirmed,
it would open up further avenues of research:  Also including Fermions, a 
time-periodically driven, isolated quantum gas should establish a temperature 
of its own, so that it might be interesting to explore, {\em e.g.\/}, whether 
an already ultracold gas can be made even colder by applying an external 
time-periodic force. In any case, these tentative speculations clearly 
indicate that time-periodically driven quantum gases offer much more than 
mere visualizations of already known  condensed-matter phenomena; in fact, 
they require the development of new concepts for time-dependent quantum 
many-body dynamics.

\ack M.H.\ acknowledges support from the Deutsche Forschungsgemeinschaft under
grant No.\ HO 1771/6. He also thanks E.~Haller for a discussion concerning
experiments performed with non-shaken optical lattices~\citep{HallerEtAl10}.
E.A.\ and O.M.\ acknowledge support from the E.U.\ through grant No.\
225187-NAMEQUAM, and from MIUR through PRIN2009. They also thank 
C.~Sias, H.~Lignier, and A.~Zenesini for their assistance.

% \appendix

% References
% ==========

% Bibliographic references with the natbib package:
% Parenthetical: \citep{Bai92} produces (Bailyn 1992).
% Textual: \citet{Bai95} produces Bailyn et al. (1995).
% An affix and part of a reference:
%   \citep[e.g.][Ch. 2]{Bar76}
%   produces (e.g. Barnes et al. 1976, Ch. 2).


\begin{thebibliography}{99}
% \bibitem[Names(Year)]{label} or \bibitem[Names(Year)Long names]{label}.
% (\harvarditem{Name}{Year}{label} is also supported.)

\bibitem[Abramowitz and Stegun(1965)]{AbramowitzStegun65}
	Abramowitz, M., Stegun, I.A. (Eds.), 1965.
	Handbook of mathematical functions.
	Dover, New York.
	
\bibitem[Alberti et al.(2009)]{AlbertiEtAl09}
	Alberti, A., Ivanov, V.V., Tino, G.M., Ferrari, G., 2009.
	Engineering the quantum transport of atomic wavefunctions over
	macroscopic distances.
	Nature Physics 5, 547--550.	

\bibitem[Arlinghaus and Holthaus(2010)]{ArlinghausHolthaus10}
	Arlinghaus, S., Holthaus, M., 2010.
	Driven optical lattices as strong-field simulators.
	Phys.\ Rev.\ A 81, 063612 [4 pages].
	
\bibitem[Arlinghaus and Holthaus(2011)]{ArlinghausHolthaus11}
	Arlinghaus, S., Holthaus, M., 2011.
	Generalized acceleration theorem for spatiotemporal Bloch waves.
	Phys.\ Rev.\ B 84, 054301 [11 pages].

\bibitem[Arlinghaus et al.(2011)]{ArlinghausEtAl11}
	Arlinghaus, S., Langemeyer, M., Holthaus, M., 2011.
	Dynamic localization in optical lattices.
	In: Keshavamurthy, S., Schlagheck, P. (Eds.),
	Dynamical Tunneling --- Theory and Experiment.
	Taylor and Francis CRC, pp.~289--310.
			
\bibitem[Ben Dahan et al.(1996)]{BenDahanEtAl96}
	Ben Dahan, M., Peik, E., Reichel, J., Castin, Y., Salomon, C., 1996.
	Bloch oscillations of atoms in an optical potential.
    	Phys.\ Rev.\ Lett.\ 76, 4508--4511.
	
\bibitem[Bloch et al.(2008)]{BlochEtAl08}
	Bloch, I., Dalibard, J., Zwerger, W., 2008.
	Many-body physics with ultracold gases.
	Rev.\ Mod.\ Phys.\ 80, 885--964.
	
\bibitem[Boers et al.(2007)]{BoersEtAl07}
	Boers, D.J., Goedeke, B., Hinrichs, D., Holthaus, M., 2007.
	Mobility edges in bichromatic optical lattices.
	Phys. Rev. A 75, 063404 [6 pages].
		
\bibitem[Breuer and Holthaus(1991)]{BreuerHolthaus91}
	Breuer, H.P., Holthaus, M., 1991.
	A semiclassical theory of quasienergies and Floquet wave functions.
	Ann.\ Phys.\ (N.Y.) 211, 249--291.
	
\bibitem[Breuer et al.(2000)]{BreuerEtAl00}
	Breuer, H.P., Huber, W., Petruccione, F., 2000.
	Quasistationary distributions of dissipative nonlinear quantum
	oscillators in strong periodic driving fields.
	Phys.\ Rev.\ E 61, 4883--4889.
	
\bibitem[Bunimovich et al.(1991)]{BunimovichEtAl91}
	Bunimovich, L., Jauslin, H.R., Lebowitz, J.L., Pellegrinotti, A.,
	Nielaba, P., 1991.
	Diffusive energy growth in classical and quantum driven oscillators.
	J.\ Stat.\ Phys.\ 62, 793--817.	
	
\bibitem[Chen et al.(2011)]{ChenEtAl11}
	Chen, Y.-A., Nascimb\`{e}ne, S., Aidelsburger, M., Atala, M.,
	Trotzky, S., Bloch, I., 2011.
	Controlling correlated tunneling and superexchange interactions with
	ac-driven optical lattices.
	Phys.\ Rev.\ Lett.\ 107, 210405 [4 pages].	
			 	
\bibitem[Chu and Telnov(2004)]{ChuTelnov04}
	Chu, S.-I, Telnov, D.A., 2004.
	Beyond the Floquet theorem: generalized Floquet formalisms and
	quasienergy methods for atomic and molecular multiphoton processes
	in intense laser fields.	
	Phys.\ Rep.\ 390, 1--131.
	
\bibitem[Creffield and Monteiro(2006)]{CreffieldMonteiro06}
	Creffield, C.E., Monteiro, T.S., 2006.
	Tuning the Mott transition in a Bose-Einstein condensate by multiple
	photon absorption.
	Phys.\ Rev.\ Lett.\ 96, 210403 [4 pages].
		
\bibitem[Drese and Holthaus(1997)]{DreseHolthaus97}
	Drese, K., Holthaus, M., 1997.
	Exploring a metal-insulator transition with ultracold atoms in
	standing light waves?
	Phys.\ Rev.\ Lett.\ 78, 2932--2935.		
	
\bibitem[Dunlap and Kenkre(1986)]{DunlapKenkre86}
	Dunlap, D.H., Kenkre, V.M., 1986.
	Dynamic localization of a charged particle moving under the influence
	of an electric field.
	Phys.\ Rev.\ B 34, 3625--3633.
	
\bibitem[Dziarmaga(2010)]{Dziarmaga10}
	Dziarmaga, J., 2010.
	Dynamics of a quantum phase transition and relaxation to a steady
	state.
	Adv.\ Phys.\ 59, 1063--1189. 

\bibitem[Eckardt et al.(2005a)]{EckardtEtAl05a}
	Eckardt, A., Jinasundera, T., Weiss, C., Holthaus, M., 2005a.
	Analog of photon-assisted tunneling in a Bose-Einstein condensate.
	Phys.\ Rev.\ Lett.\ 95, 200401 [4 pages].
						
\bibitem[Eckardt et al.(2005b)]{EckardtEtAl05b}
	Eckardt, A., Weiss, C., Holthaus, M., 2005b.
	Superfluid-insulator transition in a periodically driven optical
	lattice.
	Phys.\ Rev.\ Lett.\ 95, 260404 [4 pages].
	
\bibitem[Eckardt and Holthaus(2007)]{EckardtHolthaus07}
	Eckardt, A., Holthaus, M., 2007.
	AC-induced superfluidity.
	EPL 80, 50004 [6 pages].
	
\bibitem[Eckardt and Holthaus(2008a)]{EckardtHolthaus08a}
	Eckardt, A., Holthaus, M., 2008a.
	Dressed matter waves.
	Journal of Physics: Conference Series 99, 012007 [14 pages]
	(Available electronically from
        {\tt http://www.iop.org/EJ/toc/1742-6596/99/1}).		
	
\bibitem[Eckardt and Holthaus(2008b)]{EckardtHolthaus08} 	
	Eckardt, A., Holthaus, M., 2008b.
	Avoided-level-crossing spectroscopy with dressed matter waves.
	Phys.\ Rev.\ Lett.\ 101, 245302 [4 pages].		
	
\bibitem[Eckardt et al.(2009)]{EckardtEtAl09}
	Eckardt, A., Holthaus, M., Lignier, H., Zenesini, A., Ciampini, D.,
	Morsch, O., Arimondo, E., 2009.
	Exploring dynamic localization with a Bose-Einstein condensate.
	Phys.\ Rev.\ A 79, 013611 [7 pages].
	
\bibitem[Eckardt et al.(2010)]{EckardtEtAl10}
	Eckardt, A., Hauke, P., Soltan-Panahi, P., Becker, C.,
	Sengstock, K., Lewenstein, M., 2010.
	Frustrated quantum antiferromagnetism with ultracold bosons in 
	optical lattices.
	EPL 89, 10010 [6 pages].	
	
\bibitem[Einstein(1924)]{Einstein24}
	Einstein, A., 1924.
	Quantentheorie des einatomigen idealen Gases.
	Sitzungsberichte der Preussischen Akademie der Wissenschaften,
	XXII.~Gesamtsitzung, 261--267.

\bibitem[Einstein(1925)]{Einstein25}
	Einstein, A., 1925.
	Quantentheorie des einatomigen idealen Gases. Zweite Abhandlung.
	Sitzungsberichte der Preussischen Akademie der Wissenschaften,
	I.~Sitzung der physikalisch-mathematischen Klasse, 3--14.

\bibitem[Fainshtein et al.(1978)]{FainshteinEtAl78}
	Fainshtein, A.G., Manakov, N.L., Rapoport, L.P., 1978.
	Some general properties of quasi-energetic spectra of quantum systems
	in classical monochromatic fields.
	J.\ Phys.\ B: Atom.\ Molec.\ Phys.\ 11, 2561--2577.
	
\bibitem[Fisher et al.(1989)]{FisherEtAl89}
	Fisher, M.P.A., Weichman, P.B., Grinstein, G., Fisher, D.S., 1989.
	Boson localization and the superfluid-insulator transition.
	Phys. Rev. B 40, 546--570.
		
\bibitem[Floquet(1883)]{Floquet83}
	Floquet, G., 1883.
	Sur les \'{e}quations diff\'{e}rentielles lin\'{e}aires \`{a}
	coefficients p\'{e}riodiques.
	Ann.\ \'{E}cole Norm.\ Sup.\ 12, 47--88.
	
\bibitem[Haller et al.(2010)]{HallerEtAl10}
	Haller, E., Hart, R., Mark, M.J., Danzl, J.G., Reichs\"ollner, L.,
	N\"agerl, H.-C., 2010.
	Inducing transport in a dissipation-free lattice with super Bloch
	oscillations.
	Phys.\ Rev.\ Lett.\ 104, 200403 [4 pages].	
	
\bibitem[Holthaus(1992)]{Holthaus92}
	Holthaus, M., 1992.
	Collapse of minibands in far-infrared irradiated superlattices.
	Phys. Rev. Lett. 69, 351--354.	

\bibitem[Houston(1940)]{Houston40}
	Houston, W.V., 1940.
	Acceleration of electrons in a crystal lattice.
	Phys.\ Rev.\ 57, 184--186.	
		
\bibitem[Howland(1992a)]{Howland92a}
	Howland, J.S., 1992a.
	Quantum stability.
	In: Schr\"odinger operators: The quantum mechanical many-body problem.
	Lecture Notes in Physics 403, 100--122.
	Springer-Verlag, New York.
	
\bibitem[Howland(1992b)]{Howland92b}	
	Howland, J.S., 1992b.
	Stability of quantum oscillators.
	J.\ Phys.\ A: Math.\ Gen.\ 25, 5177--5181.
	
\bibitem[Ivanov et al.(2008)]{IvanovEtAl08}
	Ivanov, V.V., Alberti, A., Schioppo, M., Ferrari, G., Artoni, M.,
	Chiofalo, M.L., Tino, G.M., 2008.
	Coherent delocalization of atomic wave packets in driven lattice
	potentials.
	Phys.\ Rev.\ Lett.\ 100, 043602 [4 pages].	
		
\bibitem[Jaksch et al.(1998)]{JakschEtAl98}
	Jaksch, D., Bruder, C., Cirac, J.I., Gardiner, C.W., Zoller, P., 1998.
	Cold bosonic atoms in optical lattices.
	Phys.\ Rev.\ Lett.\ 81, 3108--3111.
	
\bibitem[Jaksch and Zoller(2005)]{JakschZoller05}
	Jaksch, D., Zoller, P., 2005.
	The cold atom Hubbard toolbox.
	Ann.\ Phys.\ (N.Y.) 315, 52--79.
		
\bibitem[Ketzmerick and Wustmann(2010)]{KetzmerickWustmann10}
	Ketzmerick, R., Wustmann, W., 2010.
	Statistical mechanics of Floquet systems with regular and chaotic
	states.
	Phys.\ Rev.\ E 82, 021114 [16 pages].
	
\bibitem[Kierig et al.(2008)]{KierigEtAl08}
	Kierig, E., Schnorrberger, U., Schietinger, A., Tomkovic, J.,
	Oberthaler, M.K., 2008.
	Single-particle tunneling in strongly driven double-well potentials.
	Phys.\ Rev.\ Lett.\ 100, 190405 [4 pages].	
	
\bibitem[Kohn(2001)]{Kohn01}
	Kohn, W., 2001.
	Periodic Thermodynamics.
	J.\ Stat.\ Phys.\ 103, 417--423.
	
\bibitem[Kuchment(1993)]{Kuchment93}
	Kuchment, P., 1993.
	Floquet Theory for Partial Differential Equations.
	Birkh\"auser, Basel.
	
\bibitem[Kudo et al.(2009)]{KudoEtAl09}
	Kudo, K., Boness, T., Monteiro, T.S., 2009.
	Control of bound-pair transport by periodic driving.
	Phys.\ Rev.\ A 80, 063409 [6 pages].
	
\bibitem[Kudo and Monteiro(2011a)]{KudoMonteiro11a}
	Kudo, K., Monteiro, T.S., 2011a.
	Theoretical analysis of super-Bloch oscillations.
	Phys.\ Rev.\ A 83, 053627 [6 pages].

\bibitem[Kudo and Monteiro(2011b)]{KudoMonteiro11b}
	Kudo, K., Monteiro, T.S., 2011b.	
	Periodically-driven cold atoms: the role of the phase.
	arXiv:1008.2096v8 [10 pages].
	
\bibitem[Lignier et al.(2007)]{LignierEtAl07}
	Lignier, H., Sias, C., Ciampini, D., Singh, Y., Zenesini, A.,
	Morsch, O., Arimondo, E., 2007.
	Dynamical control of matter-wave tunneling in periodic potentials.
	Phys.\ Rev.\ Lett.\ 99, 220403 [4 pages].	
	
\bibitem[London(1938)]{London38}
	London, F., 1938.
	On the Bose-Einstein condensation.
	Phys.\ Rev.\ 54, 947--954.
	
\bibitem[Ma et al.(2011)]{MaEtAl11}
	Ma, R., Tai, M.E., Preiss, P.M., Bakr, W.S., Simon, J.,
	Greiner, M., 2011.
	Photon-assisted tunneling in a biased strongly correlated Bose gas.
	Phys.\ Rev.\ Lett.\ 107, 095301 [4 pages].
	
\bibitem[Madison et al.(1997)]{MadisonEtAl97}
	Madison, K.W., Bharucha, C.F., Morrow, P.R., Wilkinson, S.R., 
	Niu, Q., Sundaram, B., Raizen, M.G., 1997.
	Quantum transport of ultracold atoms in an accelerating optical 
	potential.
	Appl. Phys. B 65, 693--700.	
	
\bibitem[Madison et al.(1998)]{MadisonEtAl98}
	Madison, K.W., Fischer, M.C., Diener, R.B., Niu, Q., Raizen, M.G., 1998.
	Dynamical Bloch band suppression in an optical lattice.
    	Phys.\ Rev.\ Lett.\ 81, 5093--5096.
		
\bibitem[Morsch et al.(2001)]{MorschEtAl01}		
	Morsch, O., M\"uller, J.H., Cristiani, M., Ciampini, D.,
	Arimondo, E., 2001.
	Bloch oscillations and mean-field effects of Bose-Einstein condensates
	in 1D optical lattices.
	Phys.\ Rev.\ Lett.\ 87, 140402 [4 pages].
	 	
\bibitem[Morsch and Oberthaler(2006)]{MorschOberthaler06}
	Morsch, O., Oberthaler, M., 2006.
	Dynamics of Bose-Einstein condensates in optical lattices.
	Rev.\ Mod.\ Phys.\ 78, 179--215.
	
\bibitem[Niu et al.(1996)]{NiuEtAl96}
	Niu, Q., Zhao, X.-G., Georgakis, G.A., Raizen, M.G., 1996.
	Atomic Landau-Zener tunneling and Wannier-Stark ladders in optical
	potentials.
    	Phys.\ Rev.\ Lett.\ 76, 4504--4507.			
		
\bibitem[Pathria(1996)]{Pathria96}
	Pathria, R.K., 1996.
	Statistical Mechanics, Second Edition.
	Butterworth-Heinemann, Oxford.
	
\bibitem[Poletti and Kollath(2011)]{PolettiKollath11}
	Poletti, D., Kollath, C., 2011.
	Slow quench dynamics of periodically driven quantum gases.
	Phys.\ Rev.\ A 84, 013615 [9 pages].		
	
\bibitem[Polkovnikov et al.(2011)]{PolkovnikovEtAl11}
	Polkovnikov, A., Sengupta, K., Silva, A., Vengalattore, M., 2011.
	Nonequilibrium dynamics of closed interacting quantum systems.
	Rev.\ Mod.\ Phys.\ 83, 863--883.
		
\bibitem[Ritus(1967)]{Ritus67}
	Ritus, V.I., 1967.
	Shift and splitting of atomic energy levels by the field of an
	electromagnetic wave.
	Sov.\ Phys.\ JETP 24, 1041--1044.
	(Originally published 1966 in
	Zh.\ Eksp.\ Teor.\ Fiz.\ 51, 1544--1549.)
				
\bibitem[Sambe(1973)]{Sambe73}
	Sambe, H., 1973.	
	Steady states and quasienergies of a quantum-mechanical system in an
	oscillating field.	
	Phys.\ Rev.\ A 7, 2203--2213.
	
\bibitem[Schneider et al.(2009)]{SchneiderEtAl09}
	Schneider, P.-I., Grishkevich, S., Saenz, A., 2009.
	Ab initio determination of Bose-Hubbard parameters for two ultracold
	atoms in an optical lattice using a three-well potential.		
	Phys.\ Rev.\ A 80, 013404 [13 pages].	
						
\bibitem[Shirley(1965)]{Shirley65}
	Shirley, J.H., 1965.	
	Solution of the Schr\"{o}dinger equation with a Hamiltonian periodic
	in  time.
	Phys.\ Rev.\ 138, B979--B987.
	
\bibitem[Sias et al.(2008)]{SiasEtAl08}
	Sias, C., Lignier, H., Singh, Y.P., Zenesini, A., Ciampini, D.,
	Morsch, O., Arimondo, E., 2008.
	Observation of photon-assisted tunneling in optical lattices.
	Phys.\ Rev.\ Lett.\ 100, 040404 [4 pages].
		
\bibitem[Slater(1952)]{Slater52}
	Slater, J.C., 1952.
	A soluble problem in energy bands.
	Phys.\ Rev.\ 87, 807--835.
	
\bibitem[Struck et al.(2011)]{StruckEtAl11} 	
	Struck, J., \"Olschl\"ager, C., Le Targat, R., Soltan-Panahi, P.,
	Eckardt, A., Lewenstein, M., Windpassinger, P., Sengstock, K., 2011.
	Quantum simulation of frustrated classical magnetism in triangular
	optical lattices.
	Science 333, 996--999.
	
\bibitem[Tokuno and Giamarchi(2011)]{TokunoGiamarchi11}
	Tokuno, A., Giamarchi, T., 2011.
	Spectroscopy for cold atom gases in periodically phase-modulated
	optical lattices.
	Phys.\ Rev.\ Lett.\ 106, 205301 [4 pages].	
		
\bibitem[Tsuji et al.(2011)]{TsujiEtAl11}
	Tsuji, N., Oka, T., Werner, P., Aoki, H., 2011.
	Dynamical band flipping in fermionic lattice systems: An
	ac-field-driven change of the interaction from repulsive to attractive.
	Phys.\ Rev.\ Lett.\ 106, 236401 [4 pages].	
		
\bibitem[Zel'dovich(1967)]{Zeldovich67}
	Zel'dovich, Ya.B., 1967.
	The quasienergy of a quantum-mechanical system subjected to a periodic
	action.
	Sov.\ Phys.\ JETP 24, 1006--1008.
	(Originally published 1966 in
	Zh.\ Eksp.\ Teor.\ Fiz.\ 51, 1492--1495.)
	
\bibitem[Zener(1934)]{Zener34}
	Zener, C., 1934.
	A theory of the electrical breakdown of solid dielectrics.
	Proc.\ R.\ Soc.\ London A 145, 523--529.	
	
\bibitem[Zenesini et al.(2009)]{ZenesiniEtAl09}
	Zenesini, A., Lignier, H., Ciampini, D., Morsch, O., Arimondo, E.,
	2009.
	Coherent control of dressed matter waves.
	Phys.\ Rev.\ Lett.\ 102, 100403 [4 pages].
	
\bibitem[Zheng et al.(2004)]{ZhengEtAl04}
	Zheng, Y., Ko\u{s}trun, M., Javanainen, J., 2004.
	Low-acceleration instability of a Bose-Einstein condensate in an
	optical lattice.
	Phys.\ Rev.\ Lett.\ 93, 230401 [4 pages].		
		
\bibitem[Zwerger(2003)]{Zwerger03}
	Zwerger, W., 2003.
	Mott-Hubbard transition of cold atoms in optical lattices.
	J.\ Opt.\ B: Quantum Semiclass.\ Opt.\ 5, S9--S16.
	
\end{thebibliography}
\end{document}